\documentclass{article}
\usepackage{graphicx}
\usepackage{cite}

\usepackage{amsmath,amssymb,exscale}

\def\mco{\multicolumn}
\newcommand{\be}{\begin{equation}}
\newcommand{\ee}{\end{equation}}
\newcommand{\ba}{\begin{eqnarray}}
\newcommand{\ea}{\end{eqnarray}}
\def\simlt{\mathrel{\lower2.5pt\vbox{\lineskip=0pt\baselineskip=0pt
             \hbox{$<$}\hbox{$\sim$}}}}
\def\simgt{\mathrel{\lower2.5pt\vbox{\lineskip=0pt\baselineskip=0pt
             \hbox{$>$}\hbox{$\sim$}}}}

\def\Journal#1#2#3#4{{#1} {\bf #2} (#4) #3}

\def\NPB{{\em Nucl. Phys.} B}
\def\PLB{{\em Phys. Lett.}  B}
\def\PRL{\em Phys. Rev. Lett.}
\def\PRD{{\em Phys. Rev.} D}

\def\JHEP{{\em JHEP} }

\def\NJP{\em New Jour. Phys.}

\begin{document}
\author{I. Antoniadis\thanks{On leave from CPHT (UMR CNRS 7644) Ecole Polytechnique, F-91128 Palaiseau}\\
Department of Physics, CERN - Theory Division,\\
 1211 Geneva 23, Switzerland}

\title{\vspace*{-0.8cm}
\begin{flushright}
\normalsize{CERN-PH-TH/2007-201}\\ 
\end{flushright}
\vspace{1cm}
Topics on String Phenomenology\footnote{Lectures given at Les Houches 2007 Summer School `String Theory and the Real World - From particle physics to astrophysics', 2 - 27 July 2007}}
%
\maketitle    
%
These lectures present some topics of string phenomenology and contain two parts.

In the first part, I review the possibility of lowering the string scale in the TeV region, that provides a theoretical framework for solving the mass hierarchy problem and unifying all interactions. The apparent weakness of gravity can then be accounted by the existence of large internal dimensions, in the submillimeter region, and transverse to a braneworld where our universe must be confined. I review the main properties of this scenario and its implications for observations at both particle colliders, and in non-accelerator gravity experiments.

In the second part, I discuss a simple framework of toroidal string models with magnetized branes, that offers an interesting self-consistent setup for string phenomenology. I will present an algorithm for fixing the geometric parameters of the compactification, build calculable particle physics models such as a supersymmetric $SU(5)$ Grand Unified Theory with three generations of quarks and leptons, and implement low energy supersymmetry breaking with gauge mediation that can be studied directly at the string level.

\newpage
\tableofcontents
\newpage
\section{Introduction}

During the last few decades, physics beyond the Standard Model (SM) was guided
from the problem of mass hierarchy. This can be formulated as the question of
why gravity appears to us so weak compared to the other three known fundamental 
interactions corresponding to the electromagnetic, weak and strong nuclear
forces. Indeed, gravitational interactions are suppressed by a very high energy
scale, the Planck mass $M_P\sim 10^{19}$ GeV, associated to a length
$l_P\sim 10^{-35}$ m, where they are expected to become
important. In a quantum theory, the hierarchy implies a severe fine tuning of the 
fundamental parameters in more than 30 decimal places in order to keep the 
masses of elementary particles at their observed values. The reason is that 
quantum radiative corrections to all masses generated by the 
Higgs vacuum expectation value (VEV) are proportional to the ultraviolet cutoff 
which in the presence of gravity is fixed by the Planck mass. As a result, all masses 
are ``attracted" to become about $10^{16}$ times heavier than their observed values.

Besides compositeness, there are three main ideas that have been proposed
and studied extensively during the last years, corresponding to different 
approaches of dealing with the mass hierarchy problem. (1) Low energy
supersymmetry with all superparticle masses in the TeV region. Indeed, in the
limit of exact supersymmetry, quadratically divergent corrections to the Higgs
self-energy are exactly cancelled, while in the softly broken case, they are
cutoff by the supersymmetry breaking mass splittings. (2) TeV scale strings,
in which quadratic divergences are cutoff by the string scale and low
energy supersymmetry is not needed. (3) Split supersymmetry, where
scalar masses are heavy while fermions (gauginos and higgsinos) are light.
Thus, gauge coupling unification and dark matter candidate are preserved
but the mass hierarchy should be stabilized by a different way and the low energy
world appears to be fine-tuned. All these ideas
are experimentally testable at high-energy particle colliders and in
particular at LHC. Below, I discuss their implementation in string theory.

The appropriate and most convenient framework for low energy supersymmetry
and grand unification is the perturbative heterotic string. Indeed, in this theory,
gravity and gauge interactions have the same origin, as massless modes of the
closed heterotic string, and they are unified at the string scale $M_s$. As a result, 
the Planck mass $M_P$ is predicted to be proportional to $M_s$:
\be
M_P=M_s/g\, ,
\label{het}
\ee
where $g$ is the gauge coupling. In the simplest constructions all gauge 
couplings are the same at the string scale, given by the four-dimensional (4d)
string coupling, and thus no grand unified group is needed for unification.
In our conventions $\alpha_{\rm GUT}=g^2\simeq 0.04$, leading to a 
discrepancy between the string and grand unification scale 
$M_{\rm GUT}$ 
by almost two orders of magnitude. Explaining this gap introduces in general
new parameters or a new scale, and the predictive power is essentially
lost. This is the main defect of this framework, which remains though an open
and interesting possibility~\cite{Dienes:1996du}.

\begin{figure}[htb]
\begin{center}
\includegraphics[width=8cm]{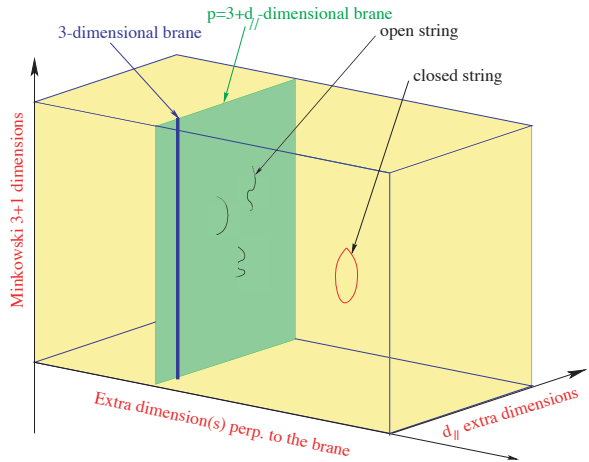}
\end{center}
\caption{\small In the type I string framework, our Universe contains,
besides the three known spatial dimensions (denoted by a single blue
line), some extra dimensions ($d_\parallel=p-3$) parallel to our world
$p$-brane (green plane) where endpoints of open strings are confined,
as well as some transverse dimensions (yellow space) where
only gravity described by closed strings can propagate.
\label{model}}
\end{figure}

The other two ideas have both as natural framework of realization type I
string theory with D-branes. Unlike in the heterotic string, gauge and gravitational
interactions have now different origin. The latter are described again by closed
strings, while the former emerge as excitations of open strings with
endpoints confined on D-branes~\cite{Angelantonj:2002ct}. 
This leads to a braneworld description of our universe, which
should be localized 
on a hypersurface, i.e. a membrane extended in $p$ spatial dimensions, 
called $p$-brane (see Fig.~\ref{model}). Closed strings propagate in all nine 
dimensions of string theory: in those extended along the $p$-brane, called parallel, 
as well as in the transverse ones. On the contrary, open strings are attached 
on the $p$-brane.
Obviously, our $p$-brane world must have at least the three known
dimensions of space. But it may contain more: the extra $d_\parallel=p-3$
parallel dimensions must have a finite size, in order to be unobservable at 
present energies, and can be as large as TeV$^{-1}\sim 10^{-18}$
m~\cite{ia}. On the other hand, transverse dimensions interact with us only 
gravitationally and experimental bounds are much weaker: their size should 
be less than about 0.1 mm~\cite{Kapner:2006si}. In the following, I review
the main properties and experimental signatures of low string 
scale models~\cite{aadd,Antoniadis:2002da}.

These lectures have two parts. In the first part, contained in sections~\ref{Framework} 
to~\ref{SMonbranes}, I describe the implementation, the properties and the main
phyical properties of low scale string theories. In the second part, contained in the
following sections, starting from section~\ref{intmagnfields}, I discuss a simple
framework of toroidal type I string compactifications with in general high string scale, 
in the presence of magnetized branes, that can be used for moduli stabilization, 
model building and supersymmetry breaking.

\section{Framework of low scale strings}\label{Framework}

In type I theory, the different origin of gauge and  gravitational interactions
implies that the relation between the Planck and string scales is not linear as
(\ref{het}) of the heterotic string. The requirement that string theory
should be weakly coupled, constrain the size of all parallel dimensions to be of
order of the string length, while transverse dimensions remain unrestricted.
Assuming an isotropic transverse space of $n=9-p$ compact 
dimensions of common radius $R_\perp$, one finds:
\begin{equation}
M_P^2=\frac{1}{g^4}M_s^{2+n}R_\perp^n\ ,\qquad
g_s \simeq g^2\, .
\label{treei}
\end{equation}
where $g_s$ is the string coupling. It follows that the type I
string scale can be chosen hierarchically smaller than the Planck 
mass~\cite{Lykken:1996fj,aadd} at
the expense of introducing extra large transverse dimensions
felt only by gravity, while keeping the string coupling small~\cite{aadd}. 
The weakness of 4d gravity compared to gauge interactions
(ratio $M_W/M_P$) is then
attributed to the largeness of the transverse space $R_\perp$
compared to the string length $l_s=M_s^{-1}$. 

\begin{figure}[htb]
\begin{center}
\includegraphics[width=4.5cm]{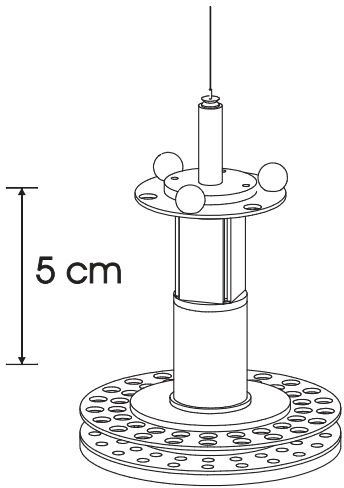}
\end{center}
\caption{\small Torsion pendulum that tested 
Newton's law at 55 $\mu$m. 
\label{newton}}
\end{figure}

An important property of these models is that gravity becomes effectively
$(4+n)$-dimensional  with a strength comparable to those of gauge
interactions at the string scale. The first relation of
Eq.~(\ref{treei}) can be understood as a consequence of the
$(4+n)$-dimensional Gauss law for gravity, with 
\be
M_*^{(4+n)}=M_s^{2+n}/g^4 
\label{GN}
\ee
the effective scale of gravity in $4+n$ dimensions.
Taking $M_s\simeq 1$ TeV,
one finds a size for the extra dimensions $R_\perp$ varying from
$10^8$ km, .1 mm, down  to a Fermi for $n=1,2$,
or 6 large dimensions, respectively. This shows that while $n=1$ is 
excluded, $n\geq 2$ is allowed by present experimental bounds
on gravitational forces~\cite{Kapner:2006si,newtonslaw}. Thus, in these models, gravity
appears to us very weak at macroscopic scales because its intensity is spread in the
``hidden" extra dimensions. At distances shorter than $R_\perp$, 
it should deviate from
Newton's law, which may be possible to explore in laboratory experiments 
(see Fig.~\ref{newton}).

The main experimental implications of TeV scale strings in particle
accelerators are of three types, in correspondence with the three different
sectors that are generally present: (i) new compactified parallel dimensions,
(ii) new extra large transverse dimensions and low scale quantum gravity, and
(iii) genuine string and quantum gravity effects. On the other hand, there
exist interesting implications in non accelerator table-top experiments
due to the exchange of gravitons or other possible states living in the bulk.

\section{Experimental implications in accelerators}
\subsection{World-brane extra dimensions}

In this case $RM_s\simgt 1$, and the associated compactification scale 
$R^{-1}_\parallel$ would be the first scale of new physics that should be 
found increasing the beam energy~\cite{ia,iab,DDG}.
There are several reasons for the existence of such dimensions.
It is a logical possibility, since out of the six extra dimensions of string theory
only two are needed for lowering the string scale, and thus the effective 
$p$-brane of our world has in general $d_\parallel\equiv p-3\le 4$.
Moreover, they can be used to address several physical problems in 
braneworld models, such as obtaining
different SM gauge couplings, explaining fermion mass hierarchies
due to different localization points of quarks and leptons in the extra 
dimensions, providing calculable mechanisms of supersymmetry breaking, etc.

The main consequence is the existence of Kaluza-Klein (KK) excitations
for all SM particles that propagate along the extra parallel dimensions.
Their masses are given by:
\be
M_m^2=M_0^2+{m^2\over R_\parallel^2}\quad ;\quad
m=0,\pm 1,\pm 2,\dots
\label{KKmass}
\ee
where we used $d_\parallel=1$, and $M_0$ is the higher dimensional mass.
The zero-mode $m=0$ is identified with the 4d state, while
the higher modes have the same quantum numbers with the lowest one, 
except for their mass given in (\ref{KKmass}).
There are two types of experimental signatures of such 
dimensions~\cite{iab,abq,AAB}:
(i) virtual exchange of KK excitations, leading to
deviations in cross-sections compared to the SM prediction, that can be used
to extract bounds on the compactification scale;
(ii) direct production of KK modes.

On general grounds, there can be two different kinds of models with
qualitatively different signatures depending on the localization properties 
of matter fermion fields. If the latter are localized in 3d
brane intersections, they do not have excitations and
KK momentum is not conserved because of the breaking of translation
invariance in the extra dimension(s). KK modes of gauge 
bosons are then singly produced giving rise to generally strong bounds on
the compactification scale and new resonances that can be observed in 
experiments. Otherwise, they can be produced only in pairs 
due to the KK momentum conservation,
making the bounds weaker but the resonances difficult to observe.

When the internal momentum is conserved, the interaction vertex involving 
KK modes has the same 4d tree-level gauge coupling.
On the other hand, their couplings to localized matter have an exponential
form factor suppressing the interactions of heavy modes.
This form factor can be viewed as the fact that the branes intersection has a
finite thickness. 
For instance, the coupling of the KK excitations of gauge fields
$A^\mu (x, y)=\sum_m
A^{\mu }_m \exp{i\frac {m y}{R_\parallel}}$ to the charge
density $j_\mu (x)$ of massless  localized fermions is described
by the effective action~\cite{ABL}: 
\be
\int d^4x \, \sum_ m e^{-\ln {16}\frac{m^2l_s^2}{2 R_\parallel^2}} \, 
j_\mu (x) \, A^{\mu }_m(x)\, .
\label{momwidth}
\ee
After Fourier transform in position space, it becomes:
\be 
\int d^4x\, dy\, 
\frac{1}{(2 \pi \ln 16)^2} e^{-\frac{y^2M_s^2}{2\ln 16}}\, 
j_\mu (x) \, A^\mu (x, y)\, ,
\label{brwidth}
\ee
from which we see that localized fermions form 
a Gaussian distribution of charge 
with a width $\sigma=\sqrt{\ln 16}\, l_s \sim 1.66 \, l_s$.

To simplify the analysis, let us consider first the case $d_\parallel=1$ 
where some of the gauge fields arise from an effective 4-brane, while 
fermions are localized states on brane intersections. 
Since the corresponding gauge couplings are reduced 
by  the size of the large dimension $R_\parallel M_s$ compared to 
the others, one can account for the ratio of the weak to strong interactions
strengths if the $SU(2)$ brane extends along the extra dimension, while
$SU(3)$ does not. 
As a result, there are 3 distinct cases to study~\cite{AAB},
denoted by $(t,l,l)$, $(t,l,t)$ and $(t,t,l)$, where the three
positions in the brackets correspond to the three SM gauge group factors
$SU(3)\times SU(2)\times U(1)$ and those with $l$ (longitudinal) feel
the extra dimension, while those with $t$ (transverse) do not.

In the $(t,l,l)$ case, there are KK excitations of $SU(2)\times U(1)$
gauge bosons: $W_{\pm}^{(m)}$, $\gamma^{(m)}$ and $Z^{(m)}$.
Performing a $\chi^2$ fit of the electroweak observables, 
one finds that if the Higgs is a bulk state $(l)$,
$R_\parallel^{-1} \simgt 3.5$ TeV~\cite{Delgado}. 
This implies that LHC can produce at most the first KK mode. 
Different choices for localization of matter and  Higgs fields lead to 
bounds, lying in the range $1-5$ TeV~\cite{Delgado}.

\begin{figure}[htb]
\begin{center}
\includegraphics[width=8cm]{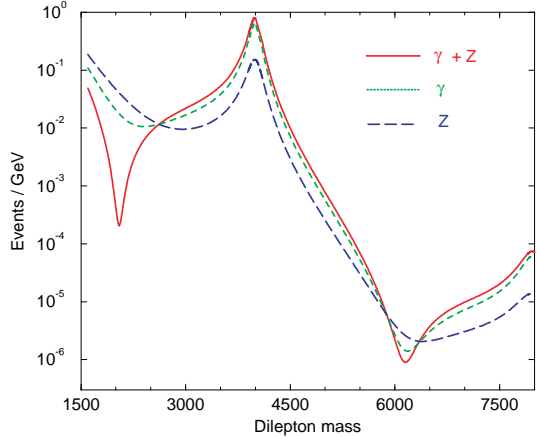}
\end{center}
\caption{\small Production of the first KK modes of the
photon and of the $Z$ boson at LHC, decaying to electron-positron pairs. 
The number of expected events is plotted
as a function of the energy of the pair in GeV. 
From highest to lowest: excitation of
$\gamma+Z$, $\gamma$ and $Z$.
\label{KK}}
\end{figure}

In addition to virtual effects, KK excitations can be produced
on-shell at LHC as new resonances~\cite{abq} (see Fig.~\ref{KK}). 
There are two different channels,
neutral Drell--Yan processes $pp \rightarrow l^+l^-X$ and the charged 
channel $l^{\pm} \nu$, corresponding to the production of the KK modes
$\gamma^{(1)},Z^{(1)}$ and $W_\pm^{(1)}$, respectively.
The discovery limits are about 6 TeV, while the exclusion bounds
15 TeV. An interesting observation in the
case of $\gamma^{(1)} + Z^{(1)}$ is that interferences can lead
to a ``dip'' just before the resonance. 
There are some ways to distinguish the corresponding signals from other
possible origin of new physics, such as models with new gauge bosons. 
In fact, in the $(t,l,l)$ and $(t,l,t)$ cases, one expects two resonances located
practically at the same mass value. This property is not shared by most of
other new gauge boson models. Moreover, the heights and widths of the
resonances are directly related to those of SM gauge bosons in
the corresponding channels.

In the $(t,l,t)$ case, only the $SU(2)$ factor feels the
extra dimension  and the limits set by the KK states of $W^{\pm}$
remain the same. On the other hand,
in the $(t,t,l)$ case where only $U(1)_Y$ feels 
the extra dimension, the limits are
weaker and the exclusion bound is around 8 TeV.
In addition to these simple possibilities, brane constructions lead often 
to cases where part of $U(1)_Y$ is $t$ and part is $l$. If
$SU(2)$ is $l$ the limits come again from $W^{\pm}$, while if it is $t$
then it will be difficult to distinguish this case from a generic extra $U(1)'$. 
A good statistics would be needed to see the deviation in the tail of the
resonance as being due to effects additional to those of a generic $U(1)'$
resonance. Finally, in the case of two or more parallel dimensions, 
the sum in the exchange of the KK modes diverges in the limit 
$R_\parallel M_s>>1$ and needs to be regularized using the form 
factor~(\ref{momwidth}). Cross-sections become bigger yielding stronger 
bounds, while resonances are closer implying that more of them 
could be reached by LHC.

On the other hand, if all SM particles propagate in the extra dimension
(called universal)\footnote{Although interesting, this scenario seems difficult
to be realized, since 4d chirality requires non-trivial action of orbifold twists 
with localized chiral states at the fixed points.}, 
KK modes can only be produced in pairs and the
lower bound on the compactification scale becomes weaker, of order of 
300-500 GeV. Moreover, no resonances can be observed at LHC, so that
this scenario appears very similar to low energy supersymmetry. In fact,
KK parity can even play the role of R-parity, implying that the lightest KK
mode is stable and can be a dark matter candidate in analogy 
to the LSP~\cite{Servant:2002aq}.

\subsection{Extra large transverse dimensions}

The main experimental signal is gravitational radiation in the bulk from
any physical process on the world-brane. In fact, the very existence of
branes breaks translation invariance in the transverse dimensions and 
gravitons can be emitted from the brane into the bulk.
During a collision of center of mass energy $\sqrt{s}$, there are 
$\sim (\sqrt{s}R_{\perp})^n$ KK excitations of gravitons with tiny masses,
that can be emitted. Each of these states 
looks from the 4d point of view as a massive, quasi-stable, 
extremely weakly coupled ($s/M^2_P$ suppressed) particle that escapes
from the detector. The total effect is a missing-energy cross-section
roughly of order: 
\be
\frac {(\sqrt{s}R_{\perp })^n} {M^2_P} \sim \frac{1}{s} 
{\left(\frac{\sqrt{s}}{M_s}\right)^{n+2}}\, .
\label{miss1}
\ee
Explicit computation of these effects leads to the bounds given in 
Table~\ref{tab:exp3}.
\begin{table}[tb]
\centering
\caption{\label{tab:exp3} Limits on $R_\perp$ in mm.} 
\vskip 0.2 cm
\small
\begin{tabular}{  | c | c | c | c |} 
\hline
  & & & \\   
{\tiny Experiment} & $n=2$ & $n=4$ & $n=6$ \\ 
\hline\hline
\mco{4}{|c|}{Collider bounds}   \\ \hline

 LEP 2   & $5\times 10^{-1}$ & $2\times 10^{-8}$  & 
                              $7 \times 10^{-11}$ \\ \hline
  Tevatron  &   $5 \times 10^{-1}$  & $10^{-8}$ 
              & $4 \times 10^{-11}$ \\ \hline 
  LHC &  $4 \times 10^{-3}$   & $6\times 10^{-10}$  & 
                              $3 \times 10^{-12}$  \\ \hline
  NLC & $ 10^{-2}$  & $10^{-9}$  & 
                              $6 \times 10^{-12}$  \\ \hline\hline
\mco{4}{|c|}{Present non-collider bounds}   \\ \hline
  
{\tiny SN1987A}   &  $3 \times 10^{-4}$   & 
           $ 10^{-8}$ 
                 & $6 \times 10^{-10} $ \\ \hline
{\tiny COMPTEL} &  $5 \times 10^{-5}$   & - & 
                              - \\ \hline
\end{tabular}
\end{table}
However, larger radii are allowed if one relaxes the assumption of isotropy, 
by taking for instance two large dimensions with different radii.

\begin{figure}[htb]
\begin{center}
\includegraphics[width=8cm]{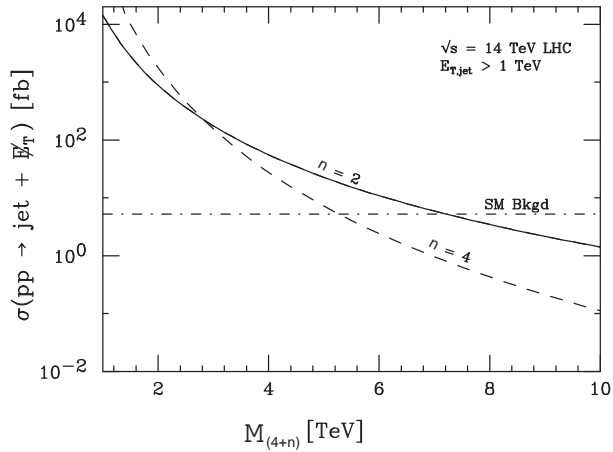}
\end{center}
\caption{\small Missing energy due to graviton emission at LHC, as a
function of the higher-dimensional gravity scale $M_*$,
produced together with a hadronic jet. The expected cross-section is shown 
for $n=2$ and $n=4$ extra dimensions, together with the SM background.
\label{graviton}}
\end{figure}

{}Fig.~\ref{graviton} shows the
cross-section for graviton emission in the bulk, corresponding to the
process $pp\to jet + graviton$ at LHC, together with the SM
background~\cite{missing}. 
{}For a given value of $M_s$, the cross-section for graviton emission
decreases with the number of large transverse dimensions, in contrast
to the case of parallel dimensions. The reason is that gravity becomes
weaker if there are more dimensions because there is more space for
the gravitational field to escape.
There is a particular energy and angular distribution  of the
produced gravitons that arise from the distribution in mass  of KK
states of spin-2. This can be contrasted to other sources of missing 
energy and might be a smoking gun for the
extra dimensional nature of such a signal.

In Table~\ref{tab:exp3}, there are also included astrophysical and cosmological
bounds. Astrophysical bounds~\cite{add2,supernovae} arise from the
requirement that the radiation of gravitons should not carry on too much
of the gravitational binding energy released during core collapse of
supernovae. In fact, the measurements of Kamiokande and IMB for SN1987A 
suggest that the main channel is neutrino fluxes.
The best cosmological bound~\cite{COMPTEL} is obtained from requiring that
decay of bulk gravitons to photons do not generate a spike in the energy
spectrum of the photon background measured by the COMPTEL instrument. Bulk 
gravitons  are  expected  to be produced just before
nucleosynthesis due to thermal radiation from the brane. The limits assume
that the temperature was at most 1 MeV as nucleosynthesis begins, and become
stronger if temperature is increased.

\subsection{String effects}

At low energies, the interaction of light (string) states is described by an
effective field theory. Their exchange generates in particular four-fermion operators
that can be used to extract independent bounds on the string scale. 
In analogy with the bounds on longitudinal extra dimensions, there are
two cases depending on the localization properties of matter fermions.
If they come from open strings with both ends on the same stack of branes,
exchange of massive open string modes gives rise to dimension eight effective
operators, involving four fermions and two space-time derivatives~\cite{Peskin,ABL}. 
The corresponding bounds on the string scale are then around 500 GeV.
On the other hand, if matter fermions are localized on non-trivial brane intersections,
one obtains dimension six four-fermion operators and the bounds become
stronger: $M_s\simgt 2-3$ TeV~\cite{ABL,Antoniadis:2002da}.
At energies higher than the string scale, new spectacular phenomena are
expected to occur, related to string physics and quantum gravity effects,
such as possible micro-black hole production~\cite{Argyres:1998qn,bh,Meade:2007sz}. 
Particle accelerators would then become the 
best tools for studying quantum gravity and string theory.

\section{Supersymmetry in the bulk and short range forces}
\subsection{Sub-millimeter forces}
 
Besides the spectacular predictions in
accelerators, there are also modifications of gravitation in
the sub-millimeter range, which can be tested in ``table-top"
experiments that measure gravity at short distances. There are three
categories of such predictions:\hfil\\  
(i) Deviations from the Newton's law $1/r^2$ behavior to $1/r^{2+n}$, 
which can be observable for $n=2$ 
large transverse dimensions of sub-millimeter size. 
This case is particularly attractive
on theoretical grounds because of the logarithmic sensitivity of SM
couplings on the size of transverse space~\cite{ab}, that allows
to determine the hierarchy~\cite{abml}.\hfil\\ 
(ii) New scalar
forces in the sub-millimeter range, related to the mechanism of
supersymmetry breaking, and mediated by light scalar fields
$\varphi$ with masses~\cite{iadd,aadd}:
\be
m_{\varphi}\simeq{m_{susy}^2\over M_P}\simeq 
10^{-4}-10^{-6}\ {\rm eV} \, ,
\label{msusy}
\ee
for a supersymmetry breaking scale $m_{susy}\simeq 1-10$ TeV. They
correspond to Compton wavelengths of 1 mm to 10 $\mu$m.
$m_{susy}$ can be either $1/R_\parallel$ if supersymmetry is broken by
compactification~\cite{iadd}, or the string scale if it is broken
``maximally" on our world-brane~\cite{aadd}. 
A universal attractive scalar force is mediated by the radion modulus
$\varphi\equiv M_P\ln R$,
with $R$ the radius of the longitudinal or transverse dimension(s).
In the former case, the result
(\ref{msusy}) follows from the behavior of the vacuum energy density
$\Lambda \sim 1/R^4_\parallel$ for large $R_\parallel$ (up to logarithmic
corrections). In the latter, supersymmetry is broken primarily on
the brane, and thus its transmission to the bulk is gravitationally
suppressed, leading to (\ref{msusy}). For $n=2$,
there may be an enhancement factor of the radion
mass by $\ln R_\perp M_s\simeq 30$ decreasing its wavelength by
an order of magnitude~\cite{abml}.

\begin{figure}[htb]
\begin{center}
\includegraphics[width=9cm]{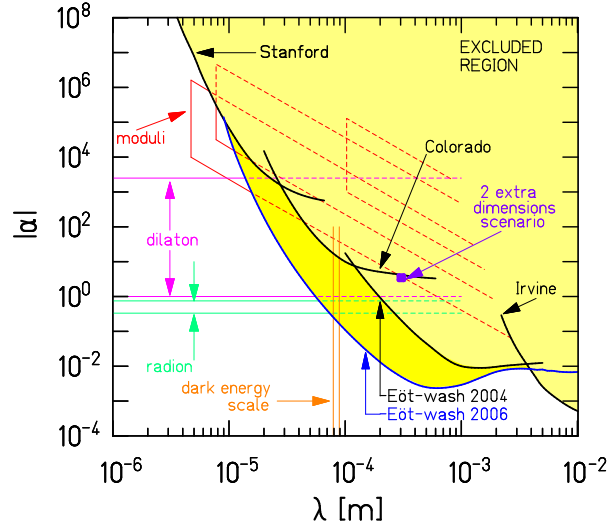}
\end{center}
\caption{\small Present limits on 
new short-range forces (yellow regions), as a function of their range $\lambda$ 
and their strength relative to gravity $\alpha$.
The limits are compared to new forces mediated by the graviton in the
case of two large extra dimensions, and by the radion.
\label{forces}}
\end{figure}

The coupling of the radius modulus 
to matter relative to
gravity can be easily computed and is given by:
\be
\sqrt{\alpha_\varphi} = {1\over M}{\partial M\over\partial\varphi}\ 
;\ \ \alpha_\varphi=\left\{ \begin{array}{l}
{\partial\ln\Lambda_{\rm QCD}\over\partial\ln R}\simeq {1\over 3}\ \ 
{\rm for}\ R_\parallel\\ \\
{2n\over n+2}=1 - 1.5\ {\rm for}\ R_\perp
\end{array}\right.
\label{dcoupling}
\ee
where $M$ denotes a generic physical mass. In the longitudinal case,
the coupling arises dominantly through the radius
dependence of the QCD gauge coupling~\cite{iadd}, while in the
case of transverse dimension, it can be deduced from the rescaling of the
metric which changes the string to the Einstein frame and depends slightly on the
bulk dimensionality ($\alpha=1-1.5$ for $n=2-6$)~\cite{abml}. 
Such a force can be tested in
microgravity experiments and should be contrasted with the change of
Newton's law due the presence of extra dimensions that is observable only
for $n=2$~\cite{Kapner:2006si,newtonslaw}.
The resulting bounds from an analysis of the radion effects
are~\cite{Adelberger:2006dh}:
\be
M_*\simgt 6\, {\rm TeV}\, .
\label{boundsradion}
\ee
In principle there can be other light moduli which couple with even
larger strengths. For example the dilaton, whose VEV determines
the string coupling, if it does not acquire
large mass from some dynamical supersymmetric mechanism, can lead to a
force of strength 2000 times bigger than gravity~\cite{tvdil}.\hfil\\
(iii) Non universal repulsive forces much stronger than gravity, mediated
by possible abelian gauge fields in the bulk~\cite{add2,akr}. Such
fields acquire tiny masses of the order of $M_s^2/M_P$, as in
(\ref{msusy}), due to brane localized anomalies~\cite{akr}. Although
their gauge coupling is infinitesimally small, $g_A\sim
M_s/M_P\simeq 10^{-16}$, it is still bigger that the gravitational
coupling $E/M_P$ for typical energies $E\sim 1$ GeV,
and the strength of the new force would be $10^6-10^8$ stronger than
gravity. This is an interesting region which will be soon
explored in micro-gravity experiments (see Fig.~\ref{forces}). Note
that in this case supernova constraints impose that there should be
at least four large extra dimensions in the
bulk~\cite{add2}.

\begin{figure}[htb]
\begin{center}
\includegraphics[width=9cm]{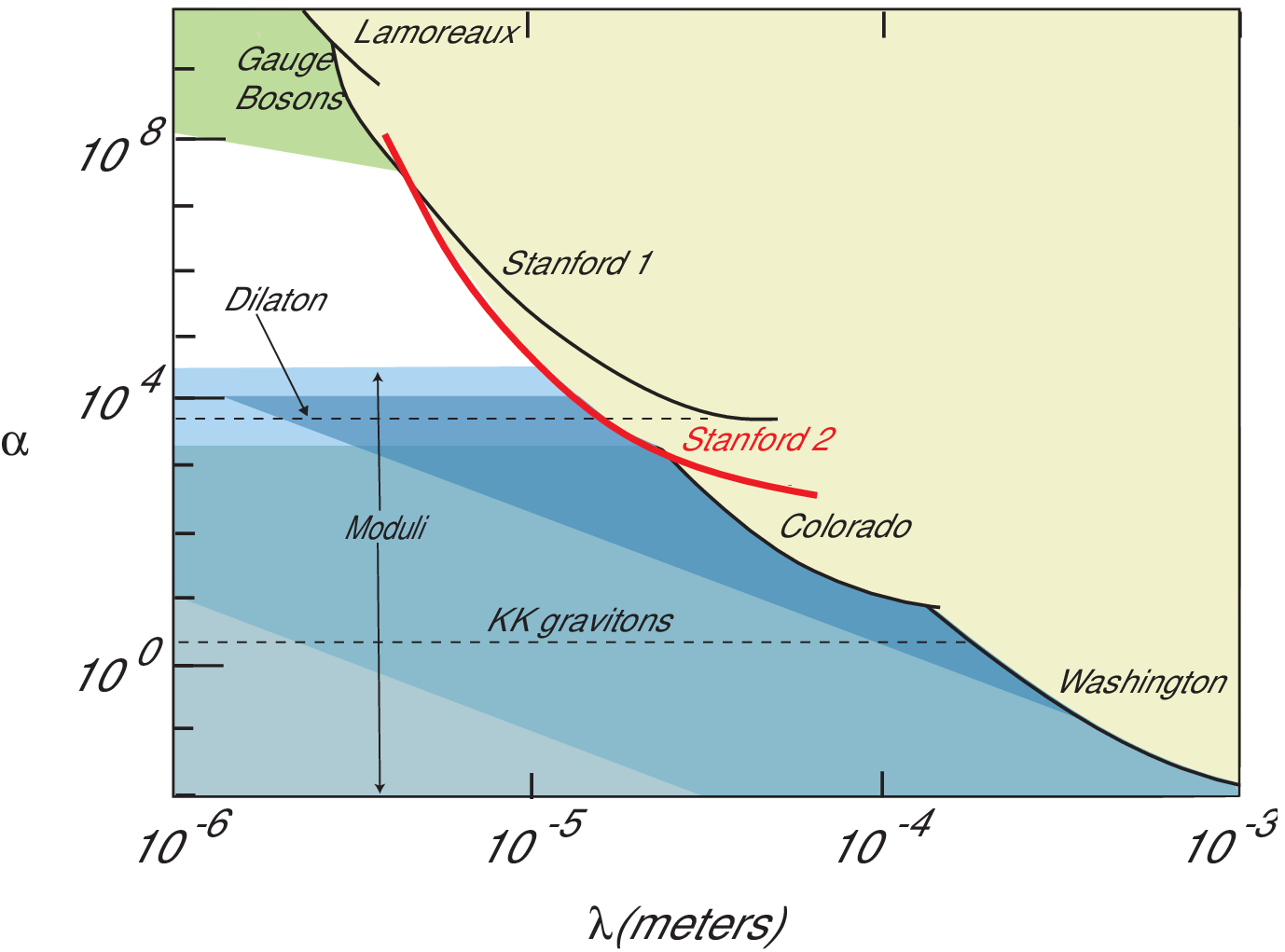}
\end{center}
\caption{\small Bounds on non-Newtonian forces in the range 6-20 $\mu$m
(see S.~J.~Smullin et al.~\cite{newtonslaw}).
\label{forcessub1}}
\end{figure}

\begin{figure}[htb]
\begin{center}
\includegraphics[width=8cm]{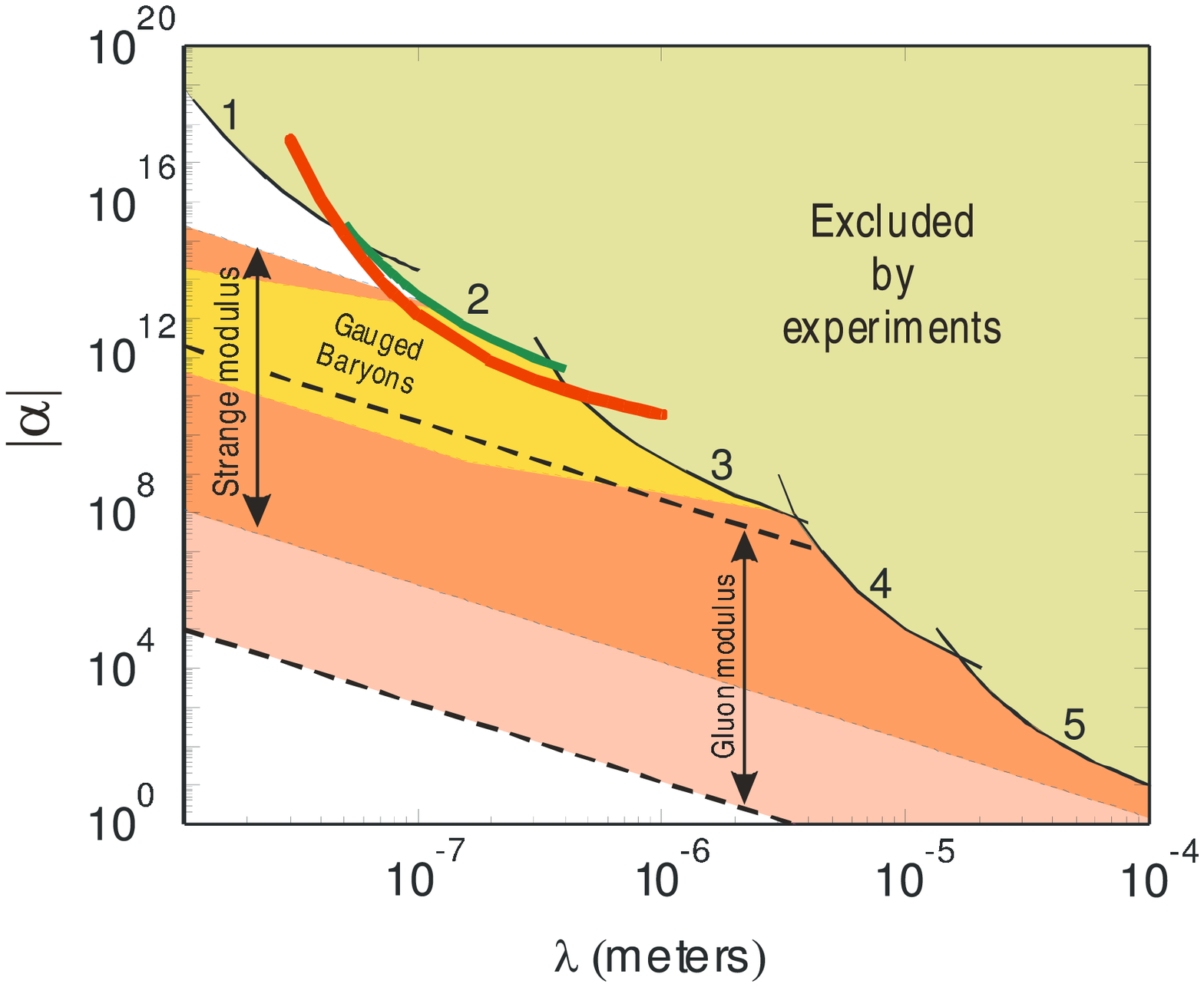}
\end{center}
\caption{\small Bounds on non-Newtonian forces in the range of 10-200 nm
(see R.~S.~Decca et al. in Ref.~\cite{newtonslaw}). Curves 4 and 5 correspond
to Stanford and Colorado experiments, respectively, of Fig.~\ref{forcessub1}
(see also J~C.~Long and J.~C.~Price of Ref.~\cite{newtonslaw}).
\label{forcessub2}}
\end{figure}

In Fig.~\ref{forces} we depict the actual information from previous,
present and upcoming experiments~\cite{newtonslaw,abml}. 
The solid lines indicate the present limits from the experiments indicated.
The excluded regions lie above these solid lines. Measuring gravitational
strength forces at short distances is challenging. 
The 
horizontal 
lines correspond to theoretical predictions, in particular for the
graviton in the case $n=2$ and for the radion in
the transverse case. These limits are compared to those
obtained from particle accelerator experiments in Table~\ref{tab:exp3}.
Finally, in Figs.~\ref{forcessub1} and \ref{forcessub2}, we display recent 
improved bounds for new forces at very short distances by focusing on the 
left hand side of Fig.~\ref{forces}, near the origin~\cite{newtonslaw}.

\subsection{Brane non-linear supersymmetry}

When the closed string sector is supersymmetric, 
supersymmetry on a generic brane configuration is non-linearly realized 
even if the spectrum is not supersymmetric and brane fields have no
superpartners. The reason is that the gravitino 
must couple to a conserved current locally, implying the existence of a 
goldstino on the brane world-volume~\cite{DM}. The goldstino is exactly massless
in the infinite (transverse) volume limit and is 
expected to acquire a small mass suppressed by the volume, 
of order (\ref{msusy}). In the standard realization, its coupling to 
matter is given via the energy momentum tensor~\cite{volaku}, 
while in general there
are more terms invariant under non-linear supersymmetry that have been
classified, up to dimension eight~\cite{bfzfer,Antoniadis:2004uk}. 

An explicit computation was performed
for a generic intersection of two brane stacks, leading to three
irreducible couplings, besides the standard 
one~\cite{Antoniadis:2004uk}: two of
dimension six involving the goldstino, a matter fermion and a scalar or
gauge field, and one four-fermion operator of dimension eight. 
Their strength is set by the goldstino decay constant $\kappa$, up to 
model-independent numerical coefficients which are 
independent of the brane angles. 
Obviously, at low energies the dominant operators are
those of dimension six. In the minimal case of (non-supersymmetric) SM,
only one of these two operators may exist, that couples the
goldstino $\chi$ with the Higgs $H$ and a lepton doublet $L$:
\be
{\cal L}_\chi^{int}=2\kappa (D_\mu H)(LD^\mu\chi)+h.c.\, ,
\label{Lchi}
\ee
where the goldstino decay constant is given by the total brane tension
\begin{equation}
{1\over 2 \ \kappa^2} = N_1 \, T_1 + N_2 \, T_2 \, ;\quad
T_i ={M_s^4 \over 4 \pi^2 g_i^2} \, ,
\label{decayconst}
\end{equation}
with $N_i$ the number of branes in each stack.
It is important to notice that the effective interaction
(\ref{Lchi}) conserves the total lepton number
$L$, as long as we assign to the goldstino a total
lepton number $L(\chi)=-1$~\cite{Antoniadis:2004se}. 
To simplify the analysis, we will consider
the simplest case where (\ref{Lchi}) exists only for the
first generation and $L$ is the electron doublet~\cite{Antoniadis:2004se}.

\begin{figure}[htb]
\begin{center}
\includegraphics[width=7.5cm]{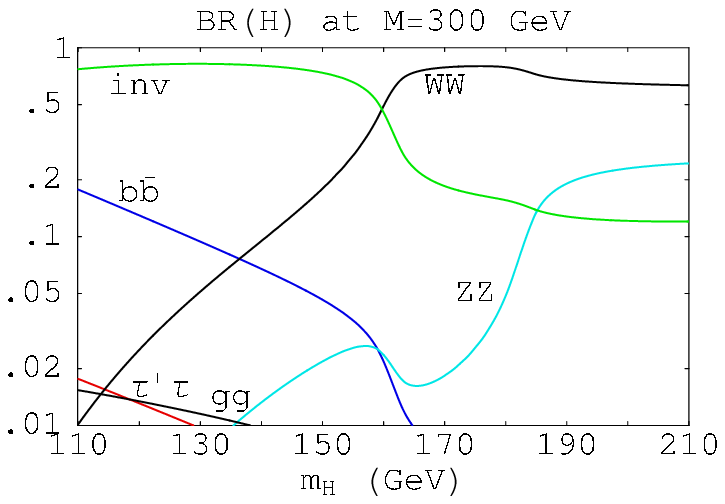}
\hskip 0.2cm
\includegraphics[width=7.5cm]{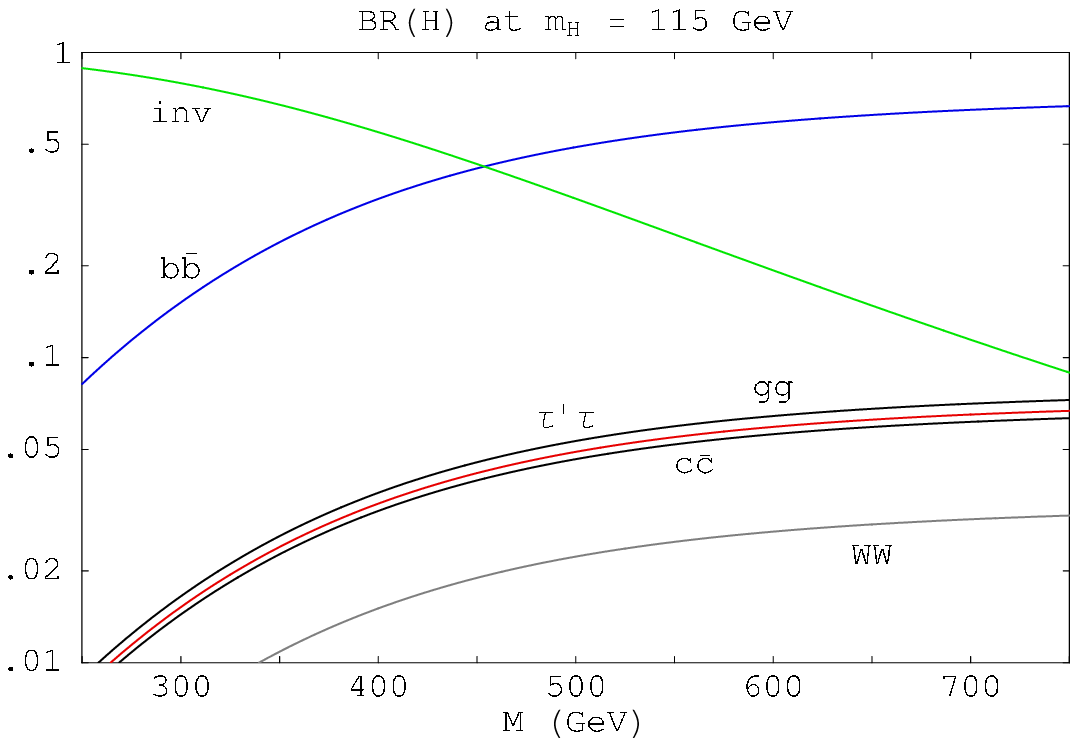}
\end{center}
\caption{\small Higgs branching rations, as functions either
of the Higgs mass $m_H$ for a fixed value of the string scale 
$M_s\simeq 2M=600$ GeV, or 
of $M\simeq M_s/2$  for $m_H=115$ GeV.
\label{Hinvisible}}
\end{figure}

The effective interaction (\ref{Lchi}) gives rise mainly to the decays
$W^\pm\to e^\pm\chi$ and $Z,H\to\nu\chi$. It turns out that the invisible
$Z$ width gives the strongest limit on $\kappa$ which can
be translated to a bound on the string scale $M_s\simgt 500$ GeV,
comparable to other collider bounds. This allows for the striking
possibility of a Higgs boson decaying dominantly, or at least with a
sizable branching ratio, via such an invisible mode, for a wide
range of the parameter space $(M_s, m_H)$, as seen in 
Fig.~\ref{Hinvisible}. 

\section{Electroweak symmetry breaking}

Non-supersymmetric  TeV strings offer also a framework to realize
gauge symmetry breaking radiatively. Indeed,
from the effective field theory point of
view, one expects quadratically divergent one-loop 
contributions to the masses of scalar fields.
The divergences are cut off by $M_s$ and if the corrections are negative,
they can induce electroweak symmetry breaking and 
explain the mild hierarchy between the weak and a string scale at a few TeV,
in terms of a loop factor~\cite{abqhiggs}.
More precisely, in the minimal case of one Higgs
doublet $H$, the scalar potential is:
\be
V=\lambda (H^\dagger H)^2 + \mu^2 (H^\dagger H)\, ,
\label{potencialh}
\ee
where $\lambda$ arises at tree-level. Moreover,
in any model where the Higgs field comes from an open string 
with both ends fixed on the same brane stack, it 
is given by an appropriate truncation of a supersymmetric theory. 
Within the minimal spectrum of the SM,
$\lambda=(g_2^2+g'^2)/8$, with $g_2$ and $g'$ the $SU(2)$ and $U(1)_Y$ gauge
couplings. On the other hand, $\mu^2$ is generated at one loop: 
\be
\label{mu2R}
\mu^2=-\varepsilon^2\, g^2\, M_s^2\, ,
\ee
where $\varepsilon$ is a loop factor that can be estimated from a
toy model computation and varies in the region 
$\epsilon\sim 10^{-1}-10^{-3}$.

Indeed, consider for illustration a simple case where the whole one-loop 
effective potential of a scalar field can be computed. We assume for 
instance one extra dimension compactified on a circle of radius $R>1$ 
(in string units). An interesting situation is provided by a class of models
where a non-vanishing VEV for a scalar (Higgs) field $\phi$ results in
shifting the mass of each KK excitation by a constant $a(\phi)$: 
\be
M^2_m= \left( \frac {m+a(\phi) }{R}\right)^2\, ,
\label{mastermass}
\ee
with $m$ the KK integer momentum number. 
Such mass shifts arise for instance in the presence 
of a Wilson line, $a= q \oint \frac{dy}{2 \pi} g A$, where $ A$ is the internal
component of a gauge field with gauge coupling $g$, and $q$ is the
charge of a given state under the corresponding generator. A
straightforward computation shows that the $\phi$-dependent part of
the one-loop effective potential is given by~\cite{Antoniadis:2001cv}:
\begin{equation}
V_{eff}=- Tr (-)^{F} \, \frac{R}{32\, \pi^{3/2} }\, \, 
\sum_{n}  e^{ 2 \pi i n a}\, \,   \int_0^\infty dl \, \,  l^{3/2} f_s(l)
\,  \,  e^{- \pi^2 n^2 R^2 l}
\label{strings}
\end{equation}
where $F =0,1$ for bosons and fermions, respectively. We have included
a regulating function $f_s(l)$ which contains for example the effects
of string oscillators.  To understand its role we will consider the
two limits $R>>1$ and $R<<1$. In the first case only the
$l\rightarrow 0$ region contributes to the integral. This means that
the effective potential receives sizable contributions only from the
infrared (field theory) degrees of freedom. In this limit we would
have $f_s(l)\rightarrow 1$.  For example, in the string model
considered in~\cite{abqhiggs}:
\be
f_s(l) = \left[\frac{1}{4 l}  \frac {\theta_2} { \eta^{3}}(il+{\frac{1}{ 2}})
\right]^4 \rightarrow 1 \qquad {\rm for }\qquad l\rightarrow 0, 
\ee
and the field theory result is finite and can be explicitly computed.
As a result of the Taylor expansion around $a=0$, we are able to
extract the one-loop contribution to the coefficient of the
term of the potential quadratic in the Higgs field. It is given by a
loop factor times the compactification scale~\cite{Antoniadis:2001cv}. 
One thus obtains $\mu^2 \sim g^2/ R^2$ up to a proportionality
constant which is calculable in the effective field theory.
On the other hand, if we consider $R \rightarrow 0$, which by
$T$-duality corresponds to taking the extra dimension as transverse
and very large, the one-loop effective potential receives
contributions from the whole tower of string oscillators as appearing
in $f_s(l)$, leading to squared masses given by a loop factor times $M_s^2$,
according to eq.~(\ref{mu2R}).

\begin{figure}[htb]
\begin{center}
\includegraphics[width=7.5cm]{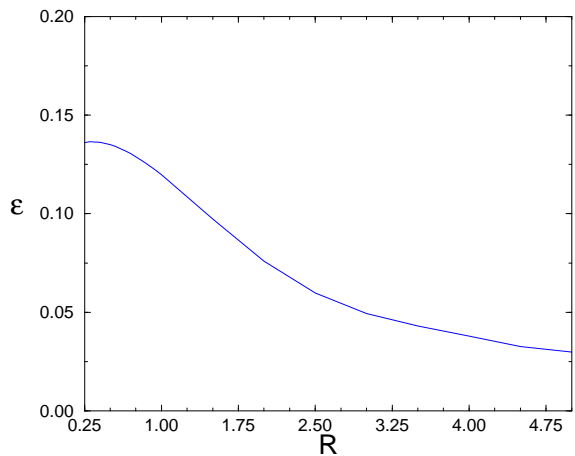}
\end{center}
\caption{\small The coefficient $\varepsilon$ of the one loop Higgs mass (\ref{mu2R}).
\label{epsilonfig}}
\end{figure}

More precisely, from the expression (\ref{strings}), one finds:
\be
\varepsilon^2(R) ={1\over 2\pi^2}\int_0^\infty \frac{dl}{\left(2\,
l\right)^{5/2}} 
{\theta_2^4\over 4\eta^{12}}\left(il+{1\over 2}\right) R^3
\sum_n n^2 e^{-2\pi n^2R^2l}\ ,
\label{epsilon2R}
\ee
which is plotted in Fig.~\ref{epsilonfig}.
For the asymptotic value $R\to 0$ (corresponding upon T-duality
to a large transverse dimension of radius $1/R$),
$\varepsilon(0)\simeq 0.14$, and the effective cut-off for the mass
term is $M_s$, as can be seen from Eq.~(\ref{mu2R}). At
large $R$, $\mu^2(R)$ falls off as $1/R^2$, 
which is the effective cut-off in the limit $R\to\infty$, as we argued above, 
in agreement with field theory
results in the presence of a compactified extra
dimension~\cite{SS, iadd}. In fact, in the
limit $R\to\infty$, an analytic approximation to $\varepsilon(R)$ gives:
\be
\varepsilon(R)\simeq \frac{\varepsilon_\infty}{M_s\, R}\, ,
\qquad\qquad
\varepsilon_\infty^2=\frac{3\, \zeta(5)}{4\, \pi^4}\simeq 0.008\, .
\label{largeR}
\ee

The potential (\ref{potencialh}) has the usual minimum, given by the
VEV of the neutral component of the Higgs doublet 
$v=\sqrt{-\mu^2/\lambda}$. Using the relation of $v$ with the $Z$
gauge boson mass, $M_Z^2=(g_2^2+g'^2)v^2/4$, and the expression
of the quartic coupling $\lambda$, 
one obtains for the Higgs mass a prediction which is the Minimal
Supersymmetric Standard Model
(MSSM) value for $\tan\beta\to\infty$ and $m_A\to\infty$:
$m_H=M_Z$. 
The tree level Higgs mass is known to receive important
radiative corrections from the top-quark sector and rises
to values around 120 GeV. 
Furthermore, from (\ref{mu2R}), one can compute $M_s$ in terms of the Higgs
mass $m_H^2=-2\mu^2$:
\begin{equation}
M_s=\frac{m_H}{\sqrt{2}\, g\varepsilon}\, ,
\label{final}
\end{equation}
yielding naturally values in the TeV range.

\section{Standard Model on D-branes}\label{SMonbranes}

The gauge group closest to the Standard Model one can easily obtain with 
D-branes is $U(3)\times U(2)\times U(1)$. The first factor arises from three
coincident ``color" D-branes. An open string with one end on
them is a triplet under $SU(3)$ and carries the same $U(1)$ charge for all three
components. Thus, the $U(1)$ factor of $U(3)$ has to be identified with {\it
gauged} baryon number. Similarly, $U(2)$ arises from two coincident ``weak"
D-branes and the corresponding abelian factor is identified with {\it gauged}
weak-doublet number. Finally, an extra $U(1)$ D-brane is necessary in 
order to accommodate the Standard Model without breaking the baryon number~\cite{akt}. 
In principle this $U(1)$ brane can be chosen to be
independent of the other two collections with its own gauge coupling. To improve
the predictability of the model, we choose to put it on top of either the
color or the weak D-branes~\cite{st}. In either case, the model has two independent gauge
couplings $g_3$ and $g_2$ corresponding, respectively, to the gauge groups $U(3)$
and $U(2)$. The $U(1)$ gauge coupling $g_1$ is equal to either $g_3$ or $g_2$.

Let us denote by $Q_3$, $Q_2$ and $Q_1$ the three $U(1)$ charges of $U(3)\times
U(2)\times U(1)$, in a self explanatory notation. Under $SU(3)\times SU(2)\times
U(1)_3\times U(1)_2\times U(1)_1$, the members of a family of quarks and
leptons have the following quantum numbers:
\ba
&Q &({\bf 3},{\bf 2};1,w,0)_{1/6}\nonumber\\
&u^c &({\bf\bar 3},{\bf 1};-1,0,x)_{-2/3}\nonumber\\
&d^c &({\bf\bar 3},{\bf 1};-1,0,y)_{1/3}\label{charges}\\
&L   &({\bf 1},{\bf 2};0,1,z)_{-1/2}\nonumber\\
&l^c &({\bf 1},{\bf 1};0,0,1)_1\nonumber
\ea
The values of the $U(1)$ charges $x,y,z,w$ will be fixed below so that
they lead to the right hypercharges, shown for completeness as subscripts.

\begin{figure}[htb]
\begin{center}
\includegraphics[width=8cm]{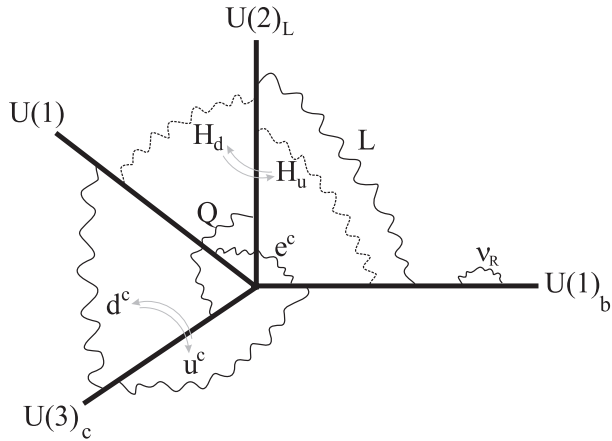}
\end{center}
\caption{\small A minimal Standard Model embedding on D-branes.
\label{SM}}
\end{figure}

It turns out that there are two possible ways of embedding the Standard Model
particle spectrum on these stacks of branes~\cite{akt}, which are shown pictorially in
Fig.~\ref{SM}.
The quark doublet $Q$ corresponds necessarily to a massless excitation of an
open string with its two ends on the two different collections of branes 
(color and weak). As seen from the figure, a fourth brane stack is needed
for a complete embedding, which is chosen to be a $U(1)_b$ extended in
the bulk. This is welcome since one can accommodate right handed neutrinos
as open string states on the bulk with sufficiently small Yukawa couplings
suppressed by the large volume of the bulk~\cite{Rnus}.
The two models are obtained by an exchange of the up and down 
antiquarks, $u^c$ and $d^c$, which correspond to open strings with one 
end on the color branes and the other either on the $U(1)$ brane, or on
the $U(1)_b$ in the bulk. The lepton doublet $L$ arises from an open string
stretched between the weak branes and $U(1)_b$, while the antilepton $l^c$
corresponds to a string with one end on the $U(1)$ brane and the other in the 
bulk. For completeness, we also show the two possible Higgs states $H_u$
and $H_d$ that are both necessary in order to give tree-level masses to
all quarks and leptons of the heaviest generation.

\subsection{Hypercharge embedding and the weak angle}

\begin{figure}[htb]
\begin{center}
\includegraphics[width=10cm]{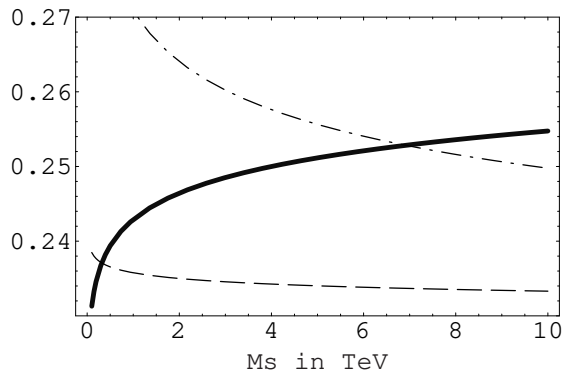}
\end{center}
\vspace{-1.0cm}
\caption{The experimental value of $\sin^2\theta_W$ (thick curve), and
the theoretical predictions (\ref{sintheta}).}
\label{sin}
\end{figure}

The weak hypercharge $Y$ is a linear combination of the three $U(1)$'s:
\be
Y= Q_1+{1\over 2}Q_2+c_3 Q_3\quad ;\quad c_3=-1/3\ {\rm or}\ 2/3\, ,
\label{Y}
\ee
where $Q_N$ denotes the $U(1)$ generator of $U(N)$ normalized so that
the fundamental representation of $SU(N)$ has unit charge. 
The corresponding $U(1)$ charges appearing in eq.~(\ref{charges}) are
$x=-1$ or $0$, $y=0$ or $1$, $z=-1$, and $w=1$ or $-1$, 
for $c_3=-1/3$ or $2/3$, respectively.
The hypercharge coupling $g_Y$ is given by~\footnote{The gauge couplings
$g_{2,3}$ are determined at the tree-level by the string coupling and other
moduli, like radii of longitudinal dimensions. In higher orders, they also
receive string threshold corrections.}:
\be
{1\over g_Y^2}={2\over g_1^2}+{4c_2^2\over g_2^2}+
{6c_3^2\over g_3^2}\, .
\label{gY}
\ee
It follows that the weak angle $\sin^2\theta_W$, is given by:
\be
\sin^2\theta_W\equiv{g_Y^2\over g_2^2+g_Y^2}=
{1\over 2+2g_2^2/g_1^2+6c_3^2g_2^2/g_3^2}\, ,
\label{sintheta}
\ee
where $g_N$ is the gauge coupling of $SU(N)$ and $g_1=g_2$ or $g_1=g_3$ 
at the string scale. In order to compare the theoretical predictions with the 
experimental value of $\sin^2\theta_W$ at $M_s$, we plot
in Fig.~\ref{sin} the corresponding curves as functions of $M_s$. 
The solid line is the experimental curve. The dashed line is the plot of the
function (\ref{sintheta}) for $g_1=g_2$ with $c_3=-1/3$ while the dotted-dashed
line corresponds to $g_1=g_3$ with $c_3=2/3$. The other two possibilities are
not shown because they lead to a value of $M_s$ which is too high to protect the
hierarchy. Thus, the second case, where the
$U(1)$ brane is on top of the color branes, is compatible with low energy data
for $M_s\sim 6-8$ TeV and $g_s\simeq 0.9$.

From Eq.~(\ref{sintheta}) and Fig.~\ref{sin}, we find the ratio of the
$SU(2)$ and $SU(3)$ gauge couplings at the string scale to be
$\alpha_2/\alpha_3\sim 0.4$. This ratio can be arranged by an appropriate
choice of the relevant moduli. For instance, one may choose the color and
$U(1)$ branes to be D3 branes while  the weak branes to be D7 branes.
Then, the ratio of couplings above can be explained by choosing the volume
of the four  compact dimensions of the seven branes to be $V_{4}=2.5$ in
string units. This being larger than one is consistent with the picture
above. Moreover it predicts an interesting spectrum of KK states for the
Standard model, different from  the naive choices that have appeared
hitherto: the only Standard Model particles that have KK descendants are
the W bosons as well as the hypercharge gauge boson. However, since the
hypercharge is a linear combination of the three $U(1)$'s, the massive $U(1)$
KK gauge bosons do not couple to the hypercharge but to the weak 
doublet number.

\subsection{The fate of $U(1)$'s, proton stability and neutrino masses}

It is easy to see that the remaining three $U(1)$ combinations orthogonal
to $Y$ are anomalous. In particular there are mixed anomalies with the
$SU(2)$ and $SU(3)$ gauge groups of the Standard Model. These anomalies are
cancelled by three axions coming from the closed string RR (Ramond) sector, via the
standard Green-Schwarz mechanism~\cite{gs}. The mixed anomalies with the
non-anomalous hypercharge are also cancelled by dimension five
Chern-Simmons type of interactions~\cite{akt}. 
An important property of the above Green-Schwarz anomaly cancellation mechanism
is that the anomalous $U(1)$ gau\-ge bosons acquire masses leaving
behind the corresponding global symmetries. This is in contrast to
what would had happened in the case of an ordinary Higgs mechanism. These global
symmetries remain exact to all orders in type I string perturbation theory
around the orientifold vacuum. This follows from the topological nature of
Chan-Paton charges in all string amplitudes. On the other hand, one expects
non-perturbative violation of global symmetries and consequently exponentially
small in the string coupling, as long as the vacuum stays at the orientifold
point. Thus, all $U(1)$ charges are conserved and since $Q_3$ is the baryon
number, proton stability is guaranteed. 

Another linear combination of the $U(1)$'s
is the lepton number. Lepton number conservation
is important for the extra dimensional neutrino mass
suppression mechanism described above, that can be destabilized by the
presence of a large Majorana neutrino mass term. Such a term can be
generated by the lepton-number violating  dimension five effective
operator $L L H H$ that leads, in the case of TeV string scale models, to
a Majorana mass of the order of a few GeV. Even if we manage to
eliminate this operator in some particular model, higher order operators
would also give unacceptably large contributions, as we focus on models
in which the ratio between the Higgs vacuum expectation value and
the string scale is just of order ${\cal O}(1/10)$. The best way to
protect tiny neutrino masses from such contributions is to impose lepton
number conservation.

A bulk neutrino propagating in $4+n$ dimensions can be decomposed
in a series of 4d KK excitations denoted collectively by $\{m\}$:
\be
S_{kin}=R_\perp^n \int d^4x\sum_{\{m\}}\left\{ {\bar\nu}_{Rm}{\slash\!\!\!\partial}\nu_{Rm}
+{\bar\nu}_{Rm}^c{\slash\!\!\!\partial}\nu_{Rm}^c+ 
{m\over R_\perp}\nu_{Rm}\nu_{Rm}^c+c.c.\right\}\, ,
\label{Skin}
\ee
where $\nu_R$ and $\nu^c_R$ are the two Weyl components of the Dirac spinor and
for simplicity we considered a common compactification radius $R_\perp$. On the other
hand, there is a localized interaction of 
$\nu_R$ with the
Higgs field and the lepton doublet, which leads to mass terms between the left-handed 
neutrino and the KK states $\nu_{Rm}$, upon the Higgs VEV $v$:
\be
S_{int}=g_s\int d^4x H(x)L(x)\nu_R(x,y=0)
\quad\rightarrow\quad {g_s v\over R_\perp^{n/2}}\sum_m\nu_L\nu_{Rm}\, ,
\label{Sint}
\ee
in strings units. Since the mass mixing $g_s v/R_\perp^{n/2}$ is much smaller than 
the KK mass $1/R_\perp$, it can be neglected for all the excitations except for the zero-mode
$\nu_{R0}$, which gets a Dirac mass with the left-handed neutrino
\be
m_\nu\simeq{g_s v\over R_\perp^{n/2}}\simeq 
v{M_s\over M_p}\simeq 10^{-3}-10^{-2}\ {\rm eV}\, ,
\label{mnu}
\ee
for $M_s\simeq 1-10$ TeV, where the relation (\ref{treei}) was used. In principle, with one bulk neutrino, one could try to explain both solar and atmospheric neutrino oscillations using also its first KK excitation. However, the later behaves like a sterile neutrino which is now excluded experimentally. Therefore, one has to introduce three bulk species (at least two) $\nu_R^i$ in order to explain neutrino oscillations in a `traditional way', using their zero-modes $\nu_{R0}^i$ \cite{LANG}. The main difference with the usual seesaw mechanism is the Dirac nature of neutrino masses, which remains an open possibility to be tested experimentally.

\section{Internal magnetic fields}\label{intmagnfields}

We now consider type I string theory, or equivalently type IIB with 
orientifold 9-planes and D9-branes~\cite{Angelantonj:2002ct}. 
Upon compactification in four dimensions on a Calabi-Yau manifold, 
one gets ${\cal N}=2$ supersymmetry in the bulk and ${\cal N}=1$ on the branes. 
We then turn on internal magnetic fields~\cite{Bachas:1995ik, Angelantonj:2000hi}, 
which, in the T-dual picture, amounts to 
intersecting branes~\cite{Berkooz:1996km, bi}. 
For generic angles, or equivalently for 
arbitrary magnetic fields, supersymmetry is spontaneously broken and 
described by effective D-terms in the four-dimensional (4d) 
theory~\cite{Bachas:1995ik}. In the weak field limit, 
$|H|\alpha'<1$ with $\alpha'$ the string Regge slope, 
the resulting mass shifts are given by:
\be
\delta M^2=(2k+1)|qH|+2qH\Sigma\quad ;\quad k=0,1,2,\dots\, ,
\label{deltam}
\ee
where $H$ is the magnetic field of an abelian 
gauge symmetry, corresponding to a  Cartan generator of the higher 
dimensional gauge group, on a non-contractible 2-cycle of the 
internal manifold. $\Sigma$ is the corresponding projection of the 
spin operator, $k$ is the Landau level and  $q=q_L+q_R$ is the charge 
of the state, given by the sum of the left and right charges of the 
endpoints of the associated open string. We recall that the exact 
string mass formula has the same form as (\ref{deltam}) with $qH$ 
replaced by:
\be
qH\longrightarrow\theta_L+\theta_R\qquad ;\qquad 
\theta_{L,R}=\arctan(q_{L,R}H\alpha')\, .
\label{stringdeltam}
\ee
Obviously, the field theory expression 
(\ref{deltam}) is reproduced in the weak field limit.

The Gauss law for the 
magnetic flux implies that the field $H$ is quantized in terms of 
the area of the corresponding 2-cycle $A$:
\be
H={m\over nA}\, ,
\label{Hquant}
\ee
where the integers $m,n$ correspond 
to the respective magnetic and electric charges; $m$ is the 
quantized flux and $n$ is the wrapping number of the higher 
dimensional brane around the corresponding internal 2-cycle.
In the T-dual representation, associated to the inversion of the compactification
radius along one of the two directions of the 2-cycle, $m$ and $n$
become the wrapping numbers around these two directions.

For simplicity, we consider first the case where 
the internal manifold is a product of three factorized tori 
$\prod_{i=1}^3 T^2_{(i)}$. Then, the mass formula (\ref{deltam})
becomes:
\be
\delta M^2=\sum_i(2k_i+1)|qH_i|+2qH_i\Sigma_i\, ,
\label{deltamI}
\ee
where $\Sigma_i$ is the projection of 
the internal helicity along the $i$-th plane. For a ten-dimensional (10d) spinor,
its eigenvalues are $\Sigma_i=\pm 1/2$, while for a 10d 
vector $\Sigma_i=\pm 1$ in one of the planes $i=i_0$
and zero in the other two $(i\ne i_0)$. Thus, charged higher dimensional
scalars become massive, fermions lead to chiral 4d zero modes if all 
$H_i\ne 0$, while the lightest scalars coming from 10d vectors have masses
\be
M_0^2=\left\{
\begin{matrix}
|qH_1|+|qH_2|-|qH_3|\cr
|qH_1|-|qH_2|+|qH_3|\cr
-|qH_1|+|qH_2|+|qH_3|\cr
\end{matrix}\, .
\right.
\label{scalars}
\ee
Note that all of them can be made positive definite, avoiding
the Nielsen-Olesen instability, if all $H_i\ne 0$. Moreover, one can 
easily show that if a scalar mass vanishes, some supersymmetry 
remains unbroken~\cite{Angelantonj:2000hi,Berkooz:1996km}.

\section{Minimal Standard Model embedding}\label{stmod}

We turn on now 
several abelian magnetic fields $H_I^a$ of different Cartan generators 
$U(1)_a$, so that the gauge group is a product of unitary factors 
$\prod_a U(N_a)$ with $U(N_a)=SU(N_a)\times U(1)_a$. In an 
appropriate T-dual 
representation, it amounts to consider several stacks of 
D6-branes intersecting in the three internal tori at angles. An 
open string with one end on the $a$-th stack has charge $\pm 1$ under 
the $U(1)_a$, depending on its orientation, and is neutral with 
respect to all others.

In this section, we perform a general study of SM embedding
in three brane stacks with gauge group $U(3)\times U(2)\times U(1)$~\cite{ar},
and present an explicit example having
realistic particle content and satisfying gauge coupling unification~\cite{Antoniadis:2004dt}.
We consider in general non oriented strings because of the presence of the
orientifold plane that gives rise mirror branes with opposite magnetic fluxes $m\to -m$
in eq.~(\ref{Hquant}). An open string stretched between a brane stack $U(N)$ and
its mirror transforms in the symmetric or antisymmetric representation, while
the multiplicity of chiral fermions is given by their intersection number.

\begin{figure}[h]
\includegraphics[height=.2\textheight]{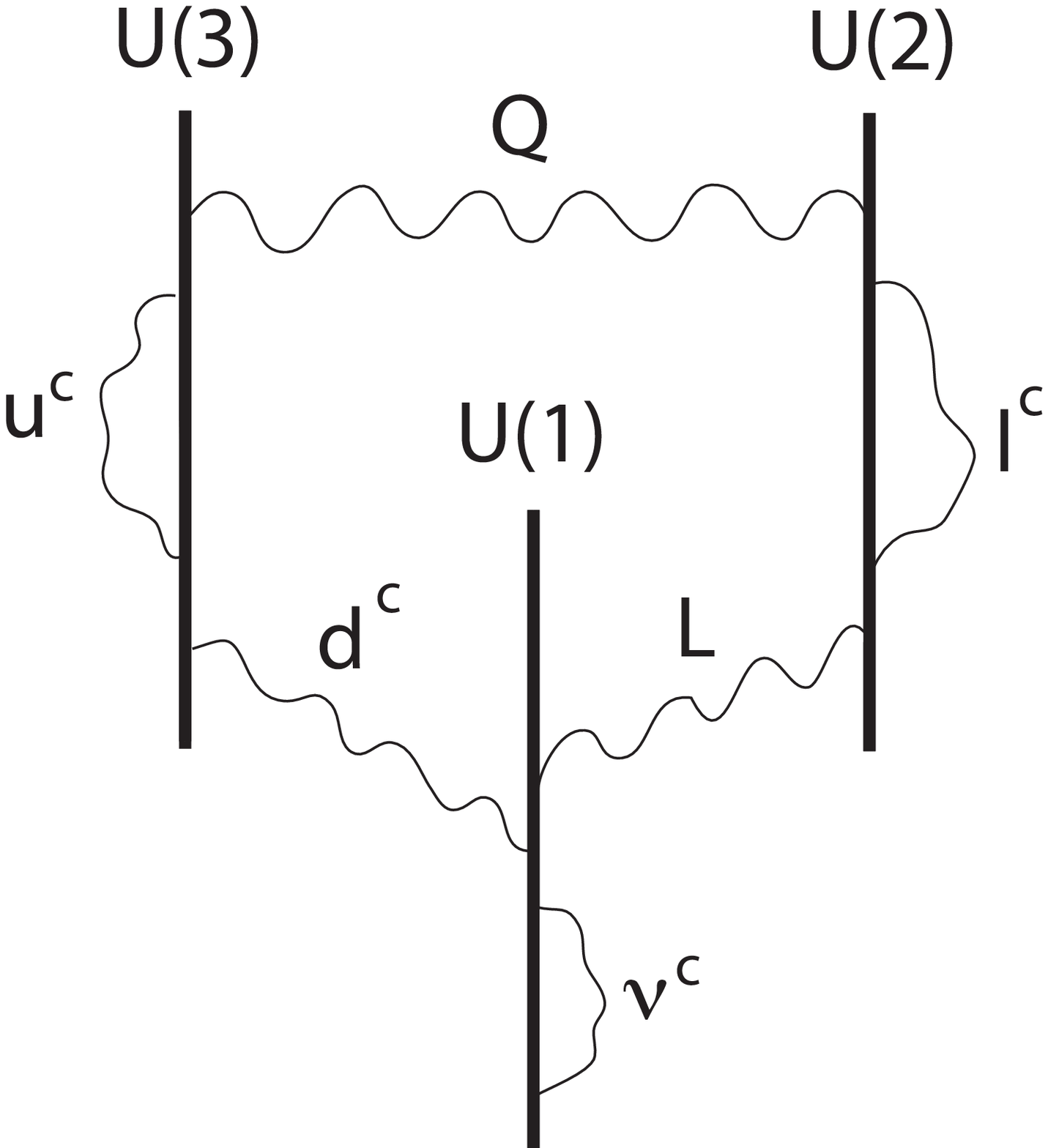}\qquad
\includegraphics[height=.2\textheight]{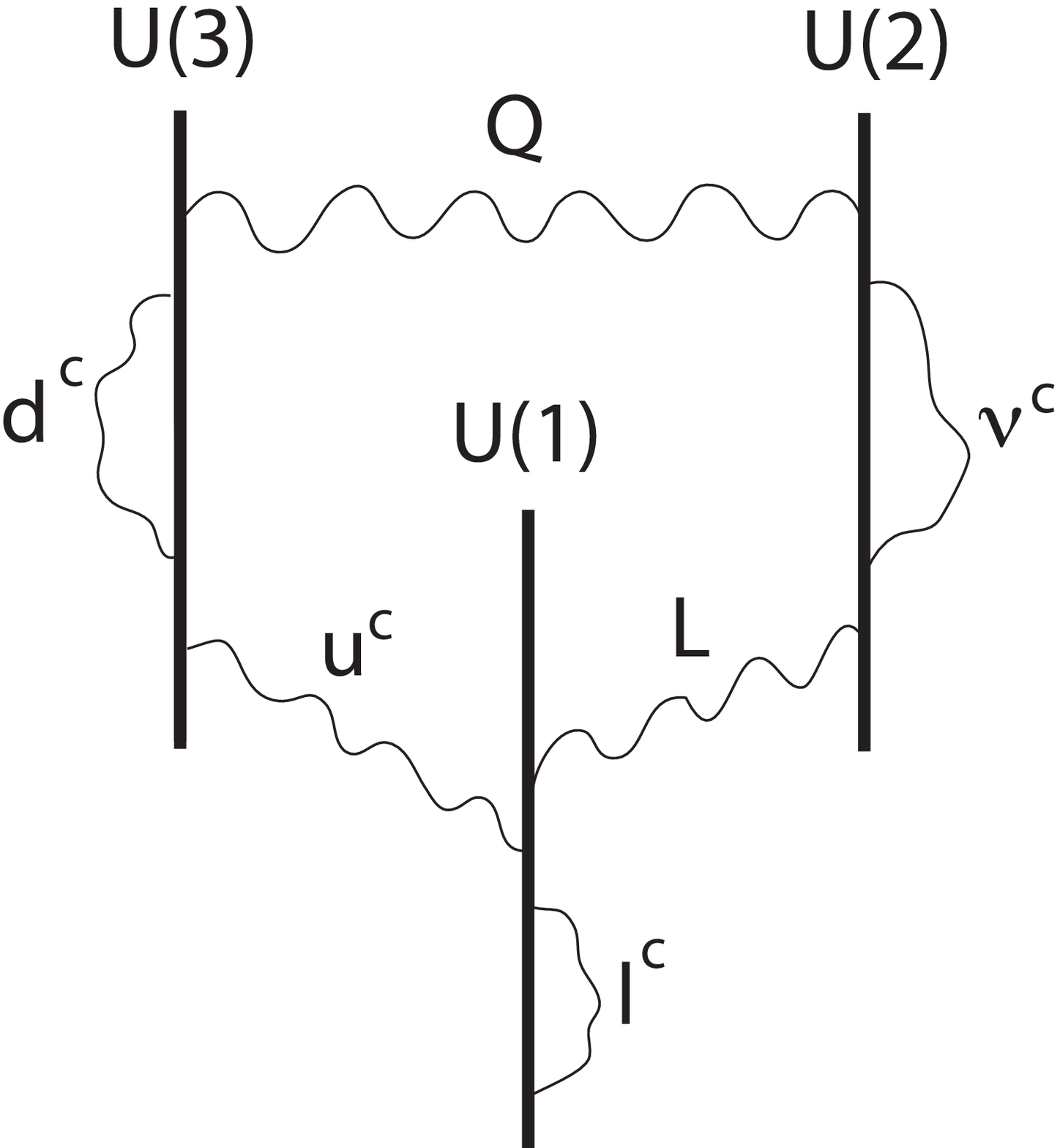}\qquad
\includegraphics[height=.2\textheight]{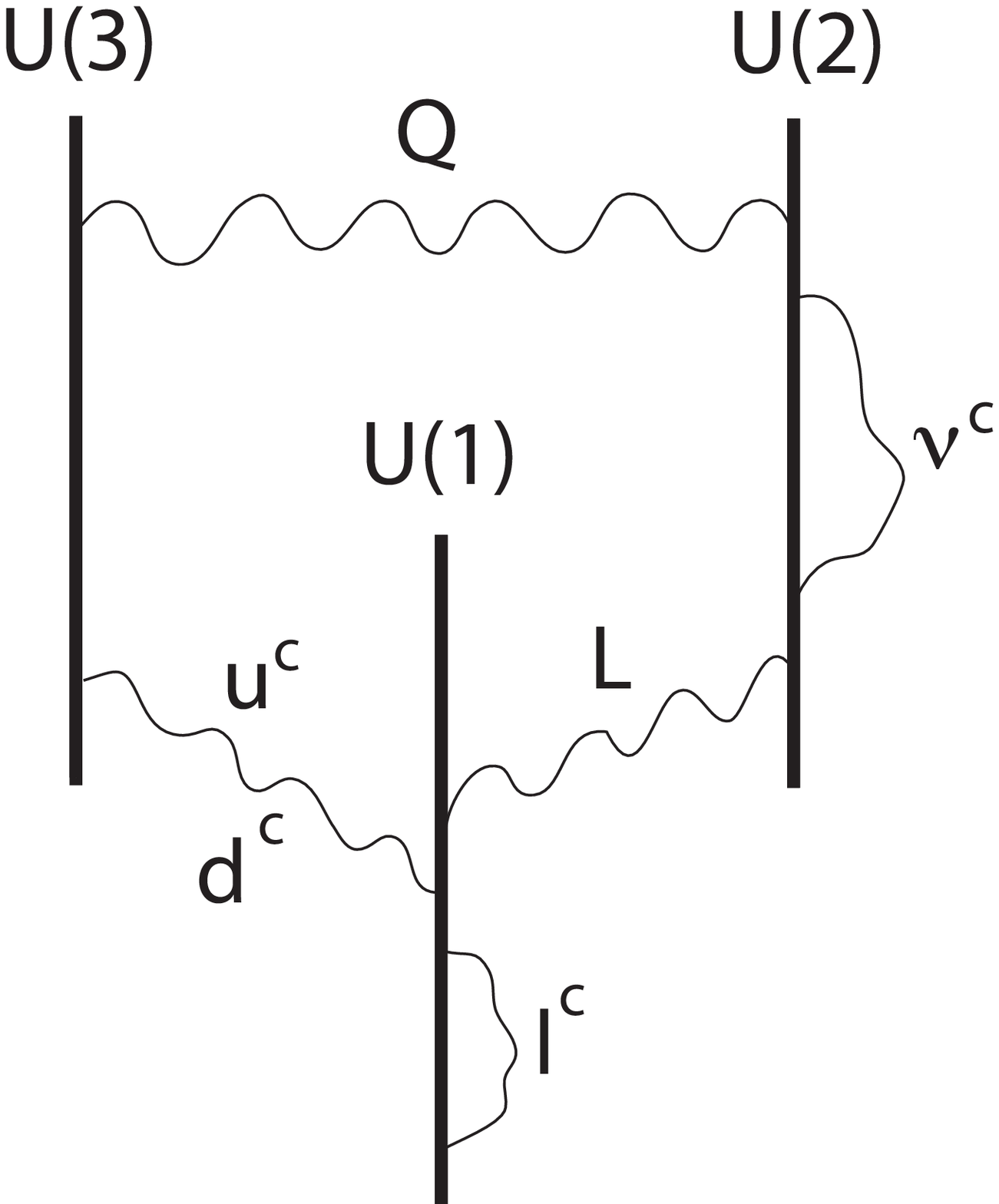}
\label{fig_modelA}
\caption{Pictorial representation of models A, B and C}
\end{figure}

The quark and lepton doublets ($Q$ and $L$) correspond 
to open strings stretched between the weak and the color or $U(1)$ 
branes, respectively. On the other hand, the $u^c$ and $d^c$ antiquarks 
can come from strings that are
either stretched between the color and $U(1)$ branes, or that have 
both ends on the color branes (stretched between the brane stack and its
orientifold image) and transform in the antisymmetric 
representation of $U(3)$ (which is an anti-triplet). There are 
therefore three possible models, depending on whether it is the $u^c$ 
(model A), or the $d^c$ (model B), or none of them (model C), the 
state coming from the antisymmetric representation of color branes. 
It follows that the antilepton $l^c$ comes in a similar way  from 
open strings with both ends either on the weak brane stack and 
transforming in the antisymmetric representation of $U(2)$ which is 
an $SU(2)$ singlet (in model A), or on the abelian brane 
and transforming in the ``symmetric" representation of $U(1)$
(in models B and C). The three models are presented pictorially 
in Fig.~\ref{fig_modelA} 


Thus, the members of a family of quarks and leptons have the 
following quantum numbers:
\ba
&&{\rm Model\ A}\qquad\qquad\quad\quad 
{\rm Model\ B}
\qquad\qquad\quad\ {\rm Model\ 
C}\nonumber\\
\!\!\!\!\!\!\!\! Q && ({\bf 3},{\bf 
2};1,1,0)_{1/6}\quad\quad\quad\ ({\bf 3},{\bf 2};1,\varepsilon_Q 
,0)_{1/6}
\qquad\, ({\bf 3},{\bf 2};1,\varepsilon_Q 
,0)_{1/6}\nonumber\\
\!\!\!\!\!\!\!\! u^c && (\bar{\bf 3},{\bf 
1};2,0,0)_{-2/3}\quad\quad\ \ \, (\bar{\bf 3},{\bf 
1};-1,0,1)_{-2/3}
\quad\ (\bar{\bf 3},{\bf 
1};-1,0,1)_{-2/3}\nonumber\\
\!\!\!\!\!\!\!\! d^c && (\bar{\bf 
3},{\bf 1};-1,0,\varepsilon_d )_{1/3}\qquad\, (\bar{\bf 3},{\bf 
1};2,0,0)_{1/3}
\qquad\ \ \, (\bar{\bf 3},{\bf 
1};-1,0,-1)_{1/3}
\label{model3}\\
\!\!\!\!\!\!\!\! L && ({\bf 1},{\bf 
2};0,-1,\varepsilon_L)_{-1/2}\quad\ \, ({\bf 1},{\bf 
2};0,\varepsilon_L,1)_{-1/2}
\quad\ \ \, ({\bf 1},{\bf 
2};0,\varepsilon_L,1)_{-1/2}\nonumber\\
\!\!\!\!\!\!\!\! l^c && ({\bf 
1},{\bf 1};0,2,0)_1\qquad\quad\quad\, ({\bf 1},{\bf 
1};0,0,-2)_1
\qquad\ \ ({\bf 1},{\bf 
1};0,0,-2)_1\nonumber\\
\!\!\!\!\!\!\!\! \nu^c && ({\bf 1},{\bf 
1};0,0,2\varepsilon_\nu)_0\qquad\quad\ ({\bf 1},{\bf 
1};0,2\varepsilon_\nu,0)_0\qquad\ \, ({\bf 1},{\bf 
1};0,2\varepsilon_\nu,0)_0\nonumber
\ea
where the last three digits after the semi-column 
in the brackets are the charges under the three 
abelian factors $U(1)_3\times U(1)_2\times U(1)$, that we will call 
$Q_3$, $Q_2$ and $Q_1$ in the following, while the subscripts denote 
the corresponding hypercharges. The various sign ambiguities 
$\varepsilon_i=\pm 1$ are due to the fact that the corresponding 
abelian factor does not participate in the hypercharge combination 
(see below). 
In the last lines, we also give the quantum numbers of a possible 
right-handed neutrino in each of the three models. These are in fact 
all possible ways of embedding the SM spectrum in three sets of 
branes.

The hypercharge combination is:
\ba
\label{hyper}
{\rm Model\ A}\quad\ &:&\quad Y=-{1\over 3}Q_3+{1\over 2}Q_2\\
{\rm Model\ B, C}&:&\quad 
Y=\ \ \, {1\over 6}Q_3-{1\over 2}Q_1\nonumber
\ea
leading to 
the following expressions for the weak angle:
\ba
{\rm Model\ A}
&:&\sin^2\theta_W={1\over 2+2\alpha_2/3\alpha_3}
={3\over 8}\, {\bigg|}_{\alpha_{_2}=\alpha_{_3}}\\
{\rm Model\ B, C}&:&
\sin^2\theta_W={1\over 1+\alpha_2/2\alpha_1+\alpha_2/6\alpha_3}\nonumber\\
& &\qquad\quad\ ={6\over 7+3\alpha_2/\alpha_1}\, 
{\bigg|}_{\alpha_{_2}=\alpha_{_3}}\nonumber
\label{sintheta_3b}
\ea
In the second part of the above equalities, we used the unification relation 
$\alpha_2=\alpha_3$, that can be 
imposed if for instance $U(3)$ and $U(2)$ branes are coincident, leading to a 
$U(5)$ unified group. 
Alternatively, this condition can be generally imposed under mild 
assumptions~\cite{Antoniadis:2004dt}.
It follows that model A admits natural gauge 
coupling unification of strong and weak interactions, 
and predicts the correct value for 
$\sin^2\theta_W=3/8$ at the unification scale $M_{\rm GUT}$.
On the other hand, model B corresponds to the flipped $SU(5)$
where the role of $u^c$ and $d^c$ is interchanged together with
$l^c$ and $\nu^c$ between the ${\bf 10}$ and ${\bf\bar 5}$ 
representations~\cite{flipped}.

Besides the hypercharge combination, there are two additional 
$U(1)$'s. It is easy to check that one of the two can be identified 
with $B-L$.  For instance, in model A choosing the signs 
$\varepsilon_d=\varepsilon_L=-\varepsilon_\nu=
-\varepsilon_H=\varepsilon_{H'}$, it is given by:
\be
B-L=-{1\over 6}Q_3+{1\over 2}Q_2-{\varepsilon_d\over 2}Q_1\, .
\label{BL}
\ee
Finally, the above spectrum can be easily implemented 
with a Higgs sector, since the Higgs field $H$ has the same 
quantum numbers as the lepton doublet or its complex conjugate. 

\section{Moduli stabilization}

Internal magnetic fluxes provide a new calculable method of moduli stabilization
 in four-dimensional (4d) type I string compactifications~\cite{AM,AKM,Bianchi:2005yz}.
In fact, moduli stabilization in the presence of 3-form closed string fluxes
led to significance progress over the last years~\cite{GKP,KKLT} but presents 
some drawbacks: (i) it has no exact string description and thus relies mainly on 
the low energy supergravity approximpation; (ii) in the generic case, it can fix 
only the complex structure and the dilaton~\cite{KST}, while for the K\"ahler class 
non-perturbative effects have to be used~\cite{KKLT}. On the other hand, 
constant internal magnetic fields can stabilize mainly K\"ahler 
moduli~\cite{Blumenhagen:2003vr,AM} and are thus complementary to 3-form 
closed string fluxes. Moreover, they can also 
be used in simple toroidal compactifications, stabilizing all geometric 
moduli in a supersymmetric vacuum using only magnetized $D9$-branes that 
have an exact perturbative string description~\cite{Fradkin:1985qd, Bachas:1995ik}.
They have also a natural implementation in intersecting D-brane models.

Here, we make use of the conventions given in Appendix A
of Ref.~\cite{AKM}, for the parametrization of the torus $T^6$, as well as
for the general definitions of the K\"ahler and complex
structure moduli. In particular, the coordinates of three
factorized tori: $(T^2)^3 \in T^6$ are given by $x_i, y_i$
$i=1,2,3$ with periodicities: $x^i = x^i + 1$, $y^i\equiv y^i + 1$, and
a volume normalization:
\be
\int dx_1\wedge dy_1\wedge dx_2\wedge dy_2\wedge dx_3
\wedge dy_3 =1\, .
\label{ft1}
\ee
The 36 moduli of $T^6$ correspond to 21 independent deformations of the
internal metric and 15 deformations of the two-index antisymmetric tensor $C_2$ from
the RR closed string sector. They form nine complex parameters of K\"ahler class
and nine of complex structure. Indeed,
the geometric  moduli decompose in a complex structure 
variation which is parametrized by the matrix $\tau_{ij}$  
entering in the definition of the complex coordinates 
\be
z_i = x_i + \tau_{ij}y^j\, ,
\label{complexbasis}
\ee
and in the K\"ahler variation of the mixed part of the metric  described by the 
real $(1,1)$-form $J = i\delta g_{i\bar{j}} dz^i \wedge d\bar{z}^j$. The later is
complexified with the corresponding RR two-form deformation.

\noindent
The stacks of $D9$-branes are characterized by three independent sets of data:
\begin{description} \vspace{-0.2cm}
\item (a) Their multiplicities $N_a$,  that describe the rank of the the 
unitary gauge group  $U(N_a)$ on each $D9$ stack. 
\item (b) The winding matrices $W_{\alpha}^{\hat{\alpha},\, a}$ describing the covering 
of the world-volume of each stack-$a$ of $D9$-branes on the compactified ambient space.
They are defined as 
$W^{\hat{\alpha}}_{\; \alpha} = {\partial \xi^{\hat{\alpha}}/\partial X^\alpha}$ for
$\alpha, \hat{\alpha} = 1,\dots, 6$,
where $\xi^{\hat{\alpha}}$ and $X^\alpha$ are the six internal coordinates on the 
world-volume and space-time, respectively. 
For simplicity, in the examples we consider here, the winding matrix $W^{\hat{\alpha}}_\alpha$ 
is chosen to be diagonal, implying that the world-volume and target space $T^6$
coordinates are identified, up to a winding multiplicity factor $n^a_{\alpha}$ 
for each brane stack-$a$:
\be
n^a_{\alpha} \equiv W_{\alpha}^{\hat{\alpha}, a}.
\label{diag-winding}
\ee
\item (c) The first Chern numbers $m^a_{\hat{\alpha} \hat{\beta}}$ of the $U(1)$ background 
on the branes world-volume. 
In other words, for each stack 
$U(N_a)=U(1)_a\times SU(N_a)$, the $U(1)_a$ has a constant field strength
on the covering of the internal space, which is a $6\times 6$ antisymmetric matrix. 
These are subject to the Dirac 
quantization condition which implies that all internal magnetic
fluxes $F^a_{\hat{\alpha}\hat{\beta}}$, on the world-volume of each stack 
of $D9$-branes, are integrally quantized. 
Explicitly, the world-volume fluxes $F^a_{\hat{\alpha}\hat{\beta}}$ and the 
corresponding target space induced fluxes  $p^a_{\alpha\beta}$
are quantized as (see (\ref{Hquant}))
\ba
\label{stab:gen:F}
 F^a_{\hat{\alpha}\hat{\beta}} &=& m^a_{\hat{\alpha}\hat{\beta}} \in \;  
\mathbb{Z} \\
p^a_{\alpha\beta} &=&  (W^{-1})_{\alpha}^{\hat{\alpha}, \; 
a} (W^{-1})_\beta^{\hat{\beta} , \; a} \,  
m^a_{\hat{\alpha}\hat{\beta}} \in \mathbb{Q}\, . \nonumber 
\ea
The complexified fluxes in the basis (\ref{complexbasis}) can be written as
\ba
F^a_{(2,0)} &\!\!\!\! =\!\!\!\! & {(\tau-\bar{\tau})^{-1}}^T \! \left[ \tau^{T} p^{a}_{xx} 
\tau - \tau^{T}{p^{a}_{xy}} - p^{a}_{yx}\tau + 
p^{a}_{yy}\right] (\tau-\bar{\tau})^{-1} \label{stab:gen:purelyholo2}\\
F^a_{(1,1)} &\!\!\!\! =\!\!\!\! & {(\tau-\bar{\tau})^{-1}}^T\! \left[ -\tau^{T} 
p^{a}_{xx}\bar{\tau} + \tau^{T}{p^{a}_{xy}} + p^{a}_{yx}\bar{\tau} - 
p^{a}_{yy}\right] (\tau-\bar{\tau})^{-1}
\label{susy-kahler0}
\ea
where the matrices $(p^{a}_{x^ix^j})$, $(p^{a}_{x^iy^j})$ and 
$(p^{a}_{y^iy^j})$
are the quantized field strengths in target space, given in eq.
(\ref{stab:gen:F}). The field strengths
 $F^a_{(2,0)}$ and $F^a_{(1,1)}$ are $3\times 3$
matrices that correspond to the upper half of the matrix 
$\mathcal{F}^a$:
\be
\mathcal{F}^a\equiv -(2\pi)^2 i\alpha' \left(
\begin{array}{cc}
F^a_{(2,0)} & F^a_{(1,1)}\\
-{F^a}^\dagger_{(1,1)} & {{F^a}^*}_{(2,0)}\\
\end{array}
\right)\, ,
\label{matrixF}
\ee
which is the total field strength in the 
cohomology basis $e_{i\bar{j}} = i dz^i \wedge d\bar{z}^j$. 
\end{description}

\subsection{Supersymmetry conditions}\label{susycond}

The supersymmetry conditions then read~\cite{AM,AKM}:
\begin{enumerate}
\item $~$\vspace{-0.4cm}
\be 
F_{(2,0)}^a = 0\qquad \forall a= 1,\dots, K\, ,
\label{susy1}
\ee
for $K$ brane stacks, stating that the purely holomorphic 
flux vanishes. For given flux quanta and winding numbers, this matrix equation 
restricts the complex structure $\tau$. Using 
eq.~(\ref{stab:gen:purelyholo2}), it imposes a restriction on the 
parameters of the complex structure matrix elements $\tau$:
\be
F_{(2,0)}^a = 0\qquad  \rightarrow \qquad \tau^{T} p^{a}_{xx} \tau - 
\tau^{T}{p^{a}_{xy}} - p^{a}_{yx}\tau + 
p^{a}_{yy} = 0\, ,
\label{stab:gen:M20_condition}
\ee
giving rise to at most six complex equations for each brane stack $a$.
\item $~$\vspace{-0.4cm}
\be
{\cal F}_a \wedge {\cal F}_a \wedge {\cal F}_a 
= {\cal F}_a \wedge J \wedge J\, ,
\label{susy2}
\ee
that gives rise to one real 
equation restricting the K\"ahler moduli. This can be understood as a 
D-flatness condition. In the 4d effective action, the 
magnetic fluxes give rise to topological couplings for the different axions of 
the compactified field theory. These arise from the dimensional reduction of the 
Wess Zumino action. In addition to the topological coupling, the ${\cal N}=1$ 
supersymmetric action yields a Fayet-Iliopoulos (FI) term of the form: 
\be
{\xi_a\over g_a^2} = {1\over (4\pi^2 \alpha')^3}
\int_{T^6}\big( {\cal F}_a \wedge {\cal F}_a\wedge {\cal F}_a -
{\cal F}_a\wedge J \wedge J\big)\, . 
\label{xioverg2}
\ee
The D-flatness condition in the  absence of charged scalars requires then 
that $\langle{\rm D}_a\rangle = \xi_a = 0$, which is equivalent to eq.~(\ref{susy2}).
In the case where $T^6$ is a product of three orthogonal 2-tori, this condition
becomes
\be
H_1+H_2+H_3=H_1H_2H_3\quad\Leftrightarrow\quad
\theta_1+\theta_2+\theta_3=0\, ,
\label{susy2factor}
\ee
in terms of the magnetic fields $H_i$ along the internal planes 
defined in section \ref{intmagnfields}, or equivalently in terms of the
angles of D6-branes with respect to the orientifold axis.
\item $~$\vspace{-0.4cm}
\be
{\rm det}W_a 
\left(J \wedge J \wedge J - {\cal F}_a \wedge {\cal F}_a \wedge J\right)> 0 \, ,
\label{susy3}
\ee
 which can also be understood from a 4d
viewpoint as the positivity of the $U(1)_a$ gauge coupling $ g_a^2$. Indeed, 
its  expression in terms of the fluxes and moduli reads
\be
{1\over g_a^2} = {1\over (4\pi^2 \alpha')^3} 
\int_{T^6}\big( J \wedge J\wedge J -{\cal F}_a\wedge {\cal F}_a \wedge J\big)\, .
\label{gaugecoupling}
\ee
\end{enumerate}

In toroidal models with NS-NS vanishing $B$-field backround, the 
net generation number of chiral fermions is in general even~\cite{RLBD}.  
Thus, it is necessary to turn on a constant $B$-field in order to obtain a
Standard Model like spectrum with three generations.  Due to the world-sheet parity
projection, the NS-NS two-index field $B_{\alpha \beta}$ is
projected out from the physical spectrum and constrained to take the discrete values
$0$ or $1/2$ (in string units) along a 2-cycle $(\alpha \beta)$ of $T^6$~\cite{Bquant}. 
Its effect is simply accounted for by shifting the target space flux matrices 
$p^a$ by $p^a+B$ in all formulae.

The main ingredients for the moduli stabilization are~\cite{AM,AKM}:
\begin{itemize}
\item
A set of nine magnetized $D9$-branes is needed to stabilize all 
36 moduli of the torus $T^6$ by the supersymmetry 
conditions~\cite{MMMS,Angelantonj:2000hi}. This follows from the second condition
(\ref{susy2}) above, in order to fix all nine K\"ahler class moduli.
At the same time, all nine corresponding $U(1)$ brane factors become 
massive by absorbing the RR partners of the K\"ahler 
moduli~\cite{Angelantonj:2000hi,AM}. This is due to a kinetic mixing between the
$U(1)$ gauge fields $A^a$ and the RR axions, arising from the 10d Chern-Simons 
coupling involving the RR two-form $C_2$ along its internal components: 
$dC_2\wedge\star(A^a\wedge\langle {\cal F}^a\rangle)$.
\item
At least six of the magnetized brane stacks must have oblique fluxes given by mutually 
non-commuting matrices, in order to fix all off-diagonal components of the metric.
The fluxes however can be chosen so that the metric is fixed in a diagonal form,
as we will see below. At the same time, the complex structure RR moduli get stabilized
by a potential generated through mixing with the metric moduli from the NS-NS 
(Neuveu-Schwarz) closed string sector~\cite{Bianchi:2005yz}.
\item
The non-linear part of Dirac-Born-Infeld 
(DBI) action which is needed to fix the overall volume. This is
only valid in 4d compactifications (and not
in higher dimensions). Indeed, in six dimensions, the condition (\ref{susy2})
becomes ${\cal F}^a\wedge J=0$ which is homogeneous in $J$ and thus
cannot fix the internal volume.
\end{itemize}

Below, we give an explicit example of nine magnetized D-brane stacks stabilizing 
all $T^6$ moduli in a way that the metric is fixed in a diagonal form~\cite{AKM}. 
The winding matrix $W^a$ is chosen to the identity, for simplicity. The first six
$U(1)$ branes with oblique fluxes are presented in Table~\ref{table1}. 
\begin{table}
\vskip -0.7cm
\caption{Six $U(1)$ branes with oblique magnetic fluxes}
\begin{center}
\vskip -0.5cm
$$
\begin{array}{|c||c|c|c|}
\hline
&&&\\
\mathrm{Stack\ \sharp} 
&\mathrm{Fluxes} & \mathrm{Fixed\ moduli} & 
\mathrm{5\! -\!brane\ localization}\\
&&&\\
\hline
\hline
&&&\\
\sharp 1  
&(F^1_{x_1 y_2},F^1_{x_2 y_1})=(1,1)& 
\tau_{31} = \tau_{32}=0 & [x_3,y_3]\\ 
N_1=1  &                               &  \tau_{11}= 
\tau_{22}  &           \\
 &                               & {\rm Re}J_{1\bar{2}}=0  &           \\
\hline
 &&&\\
\sharp 2  
& (F^2_{x_1 y_3},F^2_{x_3 y_1})=(1,1)& 
\tau_{21} = \tau_{23}=0 &  [x_2,y_2]\\
N_2=1 &                               & \tau_{11}= 
\tau_{33} & \\
 &                               & {\rm Re}J_{1\bar{3}}=0  &           \\
\hline
&&&\\
\sharp 3  
& (F^3_{x_1 x_2},F^3_{y_1 y_2})=(1,1)& 
\tau_{13}=0\, , \, \tau_{11}\tau_{22}=-1& [x_3,y_3] \\
N_3=1 
 &                          & {\rm Im}J_{1\bar{2}}=0 & \\
\hline 
&&&\\
\sharp 4  
& (F^4_{x_2 x_3},F^4_{y_2 y_3})=(1,1)& 
\tau_{12}=0  & [x_1,y_1] \\
N_4=1 
&& {\rm Im}J_{2\bar{3}}=0&\\
\hline
&&&\\
\sharp 5 
& (F^5_{x_1 x_3},F^5_{y_1 y_3})=(1,1)& {\rm Im}J_{1\bar{3}}=0 & 
 [x_2,y_2]\\
N_5=1 &&&\\
\hline
&&&\\
\sharp 6  
& (F^6_{x_2 y_3},F^6_{x_3 y_2})=(1,1)& {\rm Re}J_{2\bar{3}}=0 & 
 [x_1,y_1]\\
N_6=1 &&&\\
\hline
\end{array}
$$
\end{center}
\vskip -0.5cm
\label{table1}
\end{table}
They fix all moduli except the areas of the three factorized 2-torii.
These are fixed by adding three diagonal brane stacks displayed in the
upper part of Table~\ref{table2} (stacks $\sharp 7$, $\sharp 8$ and $\sharp 9$).
\begin{table}
\vskip -0.3cm
\caption{Brane stacks with diagonal magnetic fluxes}
\begin{center}
\vskip -0.5cm
$$
\begin{array}{|c||c|c|}
\hline
&&\\
\mathrm{Stack}\, \sharp & \mathrm{Multiplicity} &\mathrm{Fluxes} \\
&&\\
\hline\hline
&&\\
\sharp 7  & N_7=1& (F^7_{x_1 y_1},F^7_{x_2 y_2},F^7_{x_3 y_3})=(-4,-4,3)  
\\ 
\hline
 &&\\
\sharp 8  & N_8=2& (F^8_{x_1 y_1},F^8_{x_2 y_2},F^8_{x_3 y_3})=(-3,1,1)  
  \\
\hline
 && \\
\sharp 9  & N_9=3& (F^9_{x_1 y_1},F^9_{x_2 
y_2},F^9_{x_3 y_3})=(-2,3,0) \\
          \hline 
\hline
 && \\
\sharp 10  & N_{10}=2& (F^{10}_{x_1 y_1},F^{10}_{x_2 y_2},F^{10}_{x_3 y_3})=(5,1,2) \\
    \hline 
 && \\
\sharp 11  & N_{11}=2& (F^{11}_{x_1 y_1},F^{11}_{x_2 y_2},F^{11}_{x_3 y_3})=(0,4,1) \\
          \hline
            \end{array}
$$
\end{center}
\label{table2}
\vskip -0.5cm
\end{table}
These give the following restrictions on the diagonal K\"ahler moduli:
\be
\left(
\begin{array}{ccc}
{\cal F}^7_{1}&{\cal F}^7_{2}& 
{\cal F}^7_{3}\\
{\cal F}^8_{1}&{\cal F}^8_{2}& 
{\cal F}^8_{3}\\
{\cal F}^9_{1}&{\cal F}^9_{2}& 
{\cal F}^9_{3}
\end{array}
\right)
\left(
\begin{array}{c}
    J_{2}J_{3}\\
    J_{1}J_{3}\\
    J_{1}J_{2}
\end{array}
\right)
=\left(
\begin{array}{c}
    {\cal F}^7_{1}{\cal F}^7_{2}{\cal F}^7_{3}\\
    {\cal F}^8_{1}{\cal F}^8_{2}{\cal F}^8_{3}\\
    {\cal F}^9_{1}{\cal F}^9_{2}{\cal F}^9_{3}\\
\end{array}
\right)\, ,
\label{diagmod}
\ee
where we the subscript $i=1,2,3$ denotes the diagonal element $i{\bar i}$.
It follows that the moduli are fixed to the values:
\be
\tau_{ij} = i\delta_{ij}\, ; \  J_{i\bar j}=0\, ;\ 
(J_{x_1y_1},J_{x_2y_2},J_{x_3y_3}) = 4\pi^2 \alpha' \sqrt{3 \over 22}(44,66,19)\, .
\label{modfix}
\ee

Note that for every solution, an infinite discret family of vacua can be found in general 
by appropriate rescaling of fluxes and volumes. For instance, a uniform rescaling of all fluxes
by the same (integer) factor $\Lambda$ leads to new solutions where all areas $J_i$ are
rescaled by the same factor, $J_i\to \Lambda J_i$. These are large volume solutions
that remain compatible with tadpole cancellation, as we will see below.

\subsection{Tadpole cancellation conditions}

In toroidal compactifications of type I string 
theory, the magnetized $D9$-branes induce 5-brane 
charges as well, while the 
3-brane and 7-brane charges automatically vanish due to the presence of mirror 
branes with opposite flux. For general magnetic fluxes, RR
tadpole conditions can be written in 
terms of the Chern numbers and winding matrix \cite{Bianchi:2005yz,AKM}
as:
\ba
\!\!\!\!\!\! 16 &\!\!\! =\!\!\! & \sum_{a=1}^K \; N_a\;  {\rm det}W_a
\equiv \sum_{a=1}^K \;  Q^{9,\, a},
\label{tad9}
\\
\!\!\!\!\!\! 0  &\!\!\! =\!\!\! & \sum_{a=1}^K\; N_a \; {\rm det}W_a \;
\epsilon^{\alpha\beta\delta\gamma\sigma\tau} 
p^a_{\delta\gamma}p^a_{\sigma\tau}
\equiv \sum_{a=1}^K\; 
Q^{5,\, a}_{\alpha\beta}, \ \ \forall \alpha,\beta=1,\dots,6\, .
\label{tad5}
\ea
The l.h.s. of eq.~(\ref{tad9}) arises from the contribution of the 
$O9$-plane. On the other hand, in toroidal compactifications there are no 
$O5$-planes and thus  the l.h.s. of eq.~(\ref{tad5}) vanishes. 

In the example presented above, all induced 5-brane tadpoles are diagonal
despite the presence of oblique fluxes. Their localization is shown in the last
column of Table~\ref{table1}. It turns out however that the conditions
of supersymmetry and tadpole cancellation cannot be satisfied
simultaneously in toroidal compactifications, as can also be seen in our
example. Our strategy is therefore to add extra branes in order to satisfy the
RR tadpole conditions. These branes are not supersymmetric and generate a
potential for the dilaton, which is the only remaining closed string modulus 
not fixed by the supersymmetry conditions of the first nine stacks, from the FI 
D-terms (\ref{xioverg2}). One is then has two possibilites to obtain a consistent 
vacuum with stabilized moduli:

\begin{enumerate}
\item
Keep supersymmetry by turning on VEVs for charged scalars on the extra brane
stacks. In their presence, the D-flatness supersymmetry condition (\ref{susy2}) gets 
modified and in the low energy approximation, it reads: 
\begin{equation}
{\rm D}_a = - \left( \sum_\phi q^\phi_a |\phi|^2 + M_s^2\xi_a \right) =0 \, ,
\label{dterm}
\end{equation}
where $\xi_a$ is given by eqs.~(\ref{xioverg2}) and (\ref{gaugecoupling}).
The sum is extended over all scalars $\phi$ charged under the $a$-th 
$U(1)_a$ with charge $q_a^{\phi}$. When one of these scalars acquire 
a non-vanishing VEV $\langle|\phi|\rangle^2 = v_\phi^2$,  the condition (\ref{susy2}) 
is modified to:
\be
 q_a v_a^2 \int_{T^6}\big( J \wedge J\wedge J -
{\cal F}_a\wedge {\cal F}_a \wedge J\big) =
M_s^2{\int_{T^6}\big( {\cal F}_a\wedge J \wedge J-
{\cal F}_a \wedge {\cal F}_a\wedge {\cal F}_a \big)}\, .
\label{stab:gen:cond_kahlerX} 
\ee
Note that this is valid for small values
of $v_a$ (in string units), since the inclusion of charged scalars in the
D-term is in principle valid only perturbatively. 

Indeed, the model presented above can be implemented by two extra stacks
$\sharp 10$ and $\sharp 11$ with diagonal fluxes, presented in the lower 
part of Table~\ref{table2}, so that all 9- and 5-brane RR tadpoles are cancelled~\cite{AKM}.
These stacks can be made supersymmetric only in the presence of non-trivial VEV's for 
open string states charged under the corresponding $U(1)$ gauge bosons.
Let us then switch on VEV's for the fields $\phi_{10}$ and $\phi_{11}$, $v_{10}$ and  $v_{11}$
respectively, transforming in the
antisymmetric representations of the corresponding $SU(2)$ gauge groups and charged 
under the $U(1)$'s of the last two stacks. From the quanta given in 
Table~\ref{table2} and the values for the K\"ahler moduli (\ref{modfix}), the positivity 
conditions (\ref{susy3}) for these branes are satisfied. Moreover, since the 
K\"ahler form is already fixed, the supersymmetry conditions (\ref{stab:gen:cond_kahlerX}) 
determine the values of $v_{10}$ and  $v_{11}$ as:
\be
v_{10}^2 l_s^2  \simeq {0.71\over q}\simeq 0.35 \quad \quad ; 
\quad \quad  v_{11}^2 l_s^2  \simeq {0.31\over q}\simeq 0.15\, ,
\label{V:V}
\ee
where we used that the $U(1)$ charge of the fields in the antisymmetric representation is $q=2$. 
These VEV's break the two $U(1)$ factors and the final gauge group of the model becomes 
$SU(3) \times SU(2)^3$. Finally, the above values of the VEV's are reasonably small
in string units, consistently with our perturbative approach of including the charged scalar
fields in the D-terms.

We have thus presented a model where the open string moduli corresponding to charged
scalar VEV's are also fixed by the magnetic fluxes. In principle, the same method can be
applied for stabilizing other open string moduli, as well.
Note also that the discrete family
of large volume solutions is still valid for fixed $v_a$. All of them have the same
gauge symmetry but different couplings (\ref{gaugecoupling}) and matter spectra.

\item
Break supersymmetry by D-terms in a anti-de Sitter vacuum, by going ``slightly" 
off-criticality and thus generating a tree-level bulk dilaton potential 
that can also fix the dilaton at weak string coupling~\cite{ADM}.
If this breaking of supersymmetry arises on brane stacks independent from the Standard
Model, its mediation involves gauge interactions and is of particular D-type.
In particular, gauginos can
acquire Dirac masses at one loop without breaking the R-symmetry, due to the
extended supersymmetric nature of the gauge sector~\cite{Antoniadis:2006eb}. 
A more detail discussion is done in the next section.
\end{enumerate}

\subsection{Spectrum}\label{spectrum}
For completeness, here we present the spectrum of magnetized branes
in a toroidal background. The gauge sector of the spectrum follows from the open 
strings starting and ending on the same brane stack.
The gauge symmetry group is given by a  product of unitary groups 
$\otimes_a U(N_a)$, 
upon identification of the associated open strings  attached 
on a given stack  with the ones attached on its orientifold mirror. 
In addition to these vector bosons, the massless spectrum contains adjoint scalars 
and fermions forming ${\cal N}=4$, $d=4$ supermultiplets.

In the matter sector, the massless spectrum is obtained from the following 
open string states\cite{bi,Angelantonj:2000hi}:
\begin{enumerate}
\item Open strings stretched between 
the $a$-th and $b$-th stack give rise to chiral spinors in the 
bifundamental representation 
$(N_a,\bar{N}_b)$ of $U(N_a)\times U(N_b)$. Their multiplicity $I_{ab}$ is given 
by~\cite{Bianchi:2005yz}:
\be
I_{ab} = 
{ {\rm det}W_a{\rm det}W_b \over (2\pi)^3} \int_{T^6}
\left( q_a F^a_{(1, 1)} +q_b F^b_{(1, 1)}\right)^3\, ,
\label{intersection}
\ee
where $F^{a}_{(1, 1)}$ (given in eqs. (\ref{susy-kahler0}) and (\ref{matrixF}))
is the pullback of the integrally quantized world-volume flux 
$m^a_{\hat{\alpha} \hat{\beta}}$ on the target torus in the complex basis (\ref{complexbasis}), 
and $q_a$ is the corresponding $U(1)_a$ charge; in our case $q_a=+1$ $(-1)$ for the 
fundamental (anti-fundamental representation).

For factorized toroidal compactifications $T^6=(T^2)^3$
with only diagonal fluxes $p_{x^iy^i}$ $(i=1, 2, 3)$, the multiplicities of 
chiral fermions, arising from strings starting from stack $a$ and 
ending at $b$ or vice versa, take the simple form 
\be
(N_a,{\overline N}_b)  :  I_{a b} = 
\prod_i (\hat{m}_i^a \hat{n}_i^b - \hat{n}_i^a 
\hat{m}_i^b),\nonumber\\
\ee
\be
(N_a, N_b)  :  I_{ab^*} = \prod_i (\hat{m}_i^a \hat{n}_i^b + \hat{n}_i^a
\hat{m}_i^b)\, .
\label{scmult}
\ee
where the integers $\hat{m}_i^a, \hat{n}_i^a$ are defined by:
\be
\hat{m}_i^a\equiv m^a_{x^iy^i}\, ,\quad
\hat{n}^a_1 \equiv n_1^a n_2^a\, ;\quad
\hat{n}^a_2 \equiv n_3^a n_4^a\, ,\quad
\hat{n}^a_3 \equiv n_5^a n_6^a\, ,
\label{hat-m-n}
\ee
in terms of the magnetic fluxes $m^a$ and winding numbers $n^a$ of 
eqs.~(\ref{stab:gen:F}) and (\ref{diag-winding}), respectively.

\item Open strings stretched between the $a$-th brane and its 
mirror $a^\star$ give rise 
to massless modes associated to $I_{aa^\star}$ chiral fermions. 
These transform either 
in the antisymmetric or symmetric representation of $U(N_a)$. 
For factorized toroidal
compactifications $(T^2)^3$, the multiplicities of chiral fermions 
are given by;
\be
{\rm Antisymmetric}  : \quad {1\over 2}\left(\prod_i
2\hat{m}_i^a\right)\left(\prod_j \hat{n}_j^a+1\right),\nonumber \\
\ee
\be
{\rm
Symmetric}  :  \quad {1\over 2}\left(\prod_i 2\hat{m}_i^a\right)\left(\prod_j
\hat{n}_j^a-1\right).
\label{dcmult}
\ee
\end{enumerate}
In generic configurations, where supersymmetry is broken by 
the magnetic fluxes, the scalar partners of the massless chiral spinors
in twisted open string sectors ({\em i.e.} from non-trivial brane intersections)
are massive (or tachyonic).
Moreover, when a chiral index $I_{ab}$ vanishes, the corresponding intersection 
of stacks $a$ and $b$ is non-chiral. The multiplicity of the non-chiral spectrum is 
then determined by extracting the vanishing factor and calculating the 
corresponding chiral index in higher dimensions. 

\subsection{A supersymmetric $SU(5$) GUT with stabilized moduli}

A more realistic model of moduli stabilization with three generations of quarks 
an leptons can be obtained by realizing in the above framework the model A of 
section~\ref{stmod} with $U(3)$ and $U(2)$ coincident, giving rise to an $SU(5)$ 
GUT~\cite{Antoniadis:2007jq} .
To elaborate further, the model is described by twelve stacks of branes, namely 
$U_5, U_1$, $O_1,\dots , O_8$, $A$, and $B$, whose role is described below: 
\begin{itemize}
\item
The $SU(5)$ gauge group 
arises from the open string states of stack-$U_5$ containing five 
magnetized branes. The remaining eleven stacks contain only a single 
magnetized brane. Also, the stack-$U_5$ containing the GUT gauge sector, 
contributes to the GUT particle spectrum through open string states which either 
start and end on itself (or on its orientifold image) or on the stack-$U_1$, 
having only a single brane and therefore contributing an extra $U(1)$. 
More precisely, open strings stretched in the intersection 
of $U(5)$ with its orientifold image give rise to 3 chiral generations 
in the antisymmetric representation $\bf 10$ of $SU(5)$, while the 
intersection of $U(5)$ with the orientifold 
image of $U(1)$ gives 3 chiral states transforming 
as $\bf\bar 5$. Finally, the intersection of $U(5)$ with the 
$U(1)$ is non chiral, giving rise to Higgs pairs ${\bf 5}+{\bf\bar{5}}$.
The magnetic fluxes along the various branes 
are constrained by the fact that the chiral fermion spectrum, 
mentioned above, of the $SU(5)$ GUT should arise from these two sectors only. 
\item
The eight single brane stacks $O_1,\dots,O_8$, contain
oblique fluxes and generalize the set of the six stacks $\sharp 1$ - $\sharp 6$ 
of the previous toy model, in the presence of a $B$-field background needed 
to obtain odd number (three) of chiral fermions. A crucial property of these 
`oblique' branes is that the combined induced 
5-brane charge lies only along the three diagonal directions $[x_i,y_i]$.

\item
The eight `oblique' branes together with $U_5$  fix all geometric moduli by 
the supersymmetry conditions. The holomorphicity condition (\ref{susy1})
 stabilizes the complex structure moduli to the identity matrix, as in (\ref{modfix}),
 while the D-flatness condition (\ref{susy2}) for the nine stacks $U_5, O_1,\dots,O_8$,
 imposing the vanishing of the FI terms $\xi_a$ (\ref{xioverg2}), fix the nine 
 K\"ahler moduli in a diagonal form. The residual diagonal 5-brane tadpoles of the branes 
 in the stacks $U_5$, $U_1$, $O_1,\dots,O_8$ are then cancelled by introducing the 
 last two single brane stacks $A$ and $B$, satisfying also the required 9-brane charge. 
\item
 The D-flatness conditions for the brane stacks $U_1$, $A$ and $B$ can also be satisfied, 
provided  some VEVs of charged scalars living on these branes are turned on to cancel 
the corresponding FI parameters, according to eqs.~(\ref{dterm}) and
(\ref{stab:gen:cond_kahlerX}). They all take values smaller than the string scale,
consistently with their perturbative treatment, and break the three $U(1)$ symmetries.
On the other hand, the remaining nine $U(1)$ brane factors become 
massive by absorbing the RR partners of the K\"ahler class moduli. As a result,
all extra $U(1)$'s are broken and the 
only leftover gauge symmetry is an $SU(5)$ GUT. Furthermore, the 
intersections of the $U(5)$ stack with any additional brane used for 
moduli stabilization are non-chiral, yielding the three families of 
quarks and leptons in the ${\bf 10} + {\bf\bar{5}}$ representations as the only 
chiral spectrum of the model (gauge non-singlet).
\end{itemize}

\section{Gaugino masses and D-term gauge mediation}

Here, we study the possibility of breaking supersymmetry by magnetic fluxes in 
a part of the theory, instead of turning on charged scalar VEVs, such as in the brane
stacks $\sharp 10$ and $\sharp 11$ of the toy model of section~\ref{susycond}, or
in the stacks $U_1$, $A$ and $B$ of the $SU(5)$ model discussed above. 
Since this breaking of supersymmetry is induced by D-terms, gaugino masses
are vanishing at the tree-level, because they are protected by a (chiral) R-symmetry.
This symmetry is broken in general in the presence of gravity by the gravitino mass,
as well as by higher order in $\alpha'$ string corrections (on the branes).
Both effects generate gaugino masses radiatively from a diagram involving
at least one boundary, where the gauginos are localized, and having effective 
`genus' 3/2~\cite{Antoniadis:2004qn}. 

For oriented strings, there are two possibilities:
(1) one boundary and one handle, corresponding to
one gravitational loop in the effective supergravity; (2) three boundaries,
corresponding to two loops in the effective gauge theory.
In the limit of small supersymmetry breaking compared to the string scale,
both diagrams are reduced to topological amplitudes receiving contributions only
from massless states:
\begin{description} \vspace{-0.2cm}
\item (1) 
The one loop gravitational contribution of the first diagram leads to gaugino 
masses $m_{1/2}$ scaling as the third power of the gravitino mass $m_{3/2}$:
\be
m_{1/2}\propto g_s^2\frac{m_{3/2}^3}{M_s^2}\, .
\label{gravimed}
\ee
On the other hand, scalars on the brane acquire generically one-loop mass 
corrections $m_0$ from the annulus diagram~\cite{Antoniadis:1998ki}:
$m_0\simgt g_s m_{3/2}^2/M_s$,
implying that gaugino masses are suppressed relative to scalar masses:
\be
m_{1/2}^2\simlt g_s \frac{m_0^3}{M_s}\, .
\label{gravimedscalar}
\ee
Fixing $m_{1/2}$ in the TeV range, one then finds that scalars are much 
heavier $m_0\simgt 10^8$ GeV. Thus, this mechanism leads to a hierarchy between 
scalar and gaugino masses of the type required by split supersymmetry~\cite{split,splitstring}.
\item (2)
Similarly, the gauge contribution of the second diagram 
leads to even larger hierarchy:
\be
m_{1/2}\propto g_s^2 
\frac{m_0^4}{M_s^3}\, ,
\label{gaugmed}
\ee
with the proportionality constant given by the open string topological partition function 
$F_{(0,3)}$~\cite{Antoniadis:2005sd}. This result can be understood from
the supersymmetric dimension seven ciral operator in the effective field theory:
$\int d^2\theta{\cal W}^2{\rm Tr}W^2$, when the magnetized $U(1)$ gauge
superfield $\cal W$ acquires an expectation value along its D-auxiliary component:
$\langle{\cal W}\rangle=\theta\langle{\rm D}\rangle$ with $\langle{\rm D}\rangle\sim m_0^2$.
Thus, the gauginos appearing in the lowest component of the (non-ebelian) gauge superfield
$W$ acquire the Majorana mass (\ref{gaugmed}), which is in the TeV region when
the scalar masses are of order $10^{13}-10^{14}$ GeV.
\end{description}

An alternative way to generate gaugino masses is by giving Dirac type masses. Indeed,
in the toroidal models we studied above, we mentioned already that the gauge sector
on the branes comes into multiplets of ${\cal N}=4$ extended supersymmetry and thus
gauginos can be paired into Dirac massive fermions without breaking the 
R-symmetry~\cite{Antoniadis:2005em}. This leads to the possibility
of a new gauge mediation mechanism~\cite{Antoniadis:2006uj}.
A prototype model can be studied with the following setup, based on two sets of 
magnetized brane stacks: 
the observable set $\mathcal O$ and the hidden set $\mathcal
H$~\cite{Antoniadis:2005em,Antoniadis:2006eb}. 
\begin{itemize}
\item
The Standard Model  gauge sector
corresponds to open strings that propagate with both ends on the same
stack of branes that belong to $\mathcal O$: it has therefore an
extended ${\cal N}=4$ or ${\cal N}=2$ supersymmetry. Similarly, the `secluded' gauge
sector corresponds to strings with both ends on the hidden stack of
branes $\mathcal H$.
\item
The Standard Model quarks and leptons come from open strings
stretched between different stacks of branes in $\mathcal O$ that
intersect at fixed points of the internal six-torus $T^6$ and have
therefore ${\cal N}=1$ supersymmetry.
\item
The Higgs sector on the other hand corresponds to
strings stretched between different stacks of branes in $\mathcal O$
that intersect at fixed points of a $T^4$ and are parallel
along a $T^2$: it has therefore ${\cal N}=2$ supersymmetry and the two Higgs
doublets form a hypermultiplet. Finally,
the messenger sector contains strings stretched between stacks of
branes in $\mathcal O$ and the hidden branes $\mathcal H$, that
form also ${\cal N}=2$ hypermultiplets. Moreover, the two
stacks of branes along the $T^2$ are separated by a distance
$1/M$, which introduces a supersymmetric mass $M$ to the
hypermultiplet messengers. The latter are also charged under the
magnetized $U(1)$(s) that break supersymmetry in the `secluded' sector $\mathcal H$
via D-terms. 
\end{itemize}
The main properties of this mechanism are:
\begin{enumerate}
\item
The gauginos obtain Dirac masses at one loop given by:
\be
m^D_{1/2}\sim{\alpha\over 4\pi}{{\rm D}\over M}\, ,
\label{mdirac}
\ee
where $\alpha$ is the corresponding gauge coupling constant.
\item
Scalar quarks and leptons acquire masses by one-loop
diagrams involving Dirac gauginos in the effective theory where
messengers have been integrated out (three-loop diagrams in the
underlying theory). Their contributions are finite and one-loop suppressed
with respect to gaugino masses~\cite{Antoniadis:1992eb,Fox:2002bu}.
\item
The tree-level Higgs potential gets modified because of its ${\cal N}=2$
structure.
\be
V=V_{\rm soft}+{1\over 8}(g^2+g'^2)(|H_1|^2-|H_2|^2)^2
 +{1\over 2}(g^2+g'^2)|H_1H_2|^2\, ,
 \label{higgspot}
 \ee
where $H_{1,2}$ are the two Higgs doublets, $g$ and $g'$ are the $SU(2)$ and
$U(1)$ couplings, and the last term is a genuine ${\cal N}=2$ contribution 
which is absent in the MSSM. It follows that the lightest Higgs behaves as in the
(non supersymmetric) Standard Model with no $\tan\beta$ dependence on its
couplings to fermions. On the other hand, the heaviest Higgs plays no role in
electroweak symmetry breaking and does not couple to the $Z$-boson. In fact,
the model behaves as the MSSM at large $\tan\beta$ and the `little' fine-tuning
problem is significantly reduced~\cite{Antoniadis:2006eb}.
 \item
The supersymmetric flavor problem is solved as in usual gauge mediation. 
Moreover, there is a common supersymmetry breaking scale in the observable
sector, the masses of all supersymmetric particles being proportional
to powers of gauge couplings. Finally, there are distinct collider signals
different from that of the MSSM.
\end{enumerate}

In conclusion, the framework of toroidal string compactifications with magnetized 
branes described above, starting from section~\ref{intmagnfields},
offers an interesting self-consistent setup for string
phenomenology, in which one can build simple calculable models of particle
physics with stabilized moduli and implement low energy supersymmetry
breaking that can be studied directly at the string level.

\section*{Acknowledgments}
\vskip 0.2cm

This work was supported in part by the European Commission under the
RTN contract MRTN-CT-2004-503369, and in part by the INTAS
contract 03-51-6346.

\end{document}